\documentclass[modern]{aastex631}
\usepackage[utf8]{inputenc}
\usepackage{amsmath}
\usepackage{longtable}
\usepackage{threeparttablex}
\usepackage{multirow}
\usepackage{graphicx}
\usepackage{upgreek}
\usepackage{enumitem}
\setitemize{noitemsep,topsep=0pt,parsep=0pt,partopsep=0pt}

\usepackage[colorinlistoftodos]{todonotes}

\usepackage{chemformula}
\usepackage{xr-hyper}
\usepackage{hyperref}
\usepackage{xcolor}
\usepackage{color, colortbl}

\usepackage{soul}
	
\definecolor{Gray}{gray}{0.92}
\definecolor{turquoisehl}{rgb}{0,1,1}
\definecolor{redhl}{rgb}{1,0.5,0.5}
\definecolor{greenhl}{rgb}{0.5,1.0,0.5}
\definecolor{pinkhl}{rgb}{0.9,0.4,0.9}

\def\fO2{\textit{f}\ch{O2}}

\makeatletter
\newcommand*{\addFileDependency}[1]{% argument=file name and extension
  \typeout{(#1)}
  \@addtofilelist{#1}
  \IfFileExists{#1}{}{\typeout{No file #1.}}
}
\makeatother

\newcommand*{\myexternaldocument}[1]{%
    \externaldocument{#1}%
    \addFileDependency{#1.tex}%
    \addFileDependency{#1.aux}%
}

\myexternaldocument{supplement}

\received{August 18, 2022}
\submitjournal{The Astrophysical Journal} 

\shorttitle{VapoRock: Modeling vaporized silicate melts}
\shortauthors{Wolf et al.}

\begin{document}

\title{VapoRock: Thermodynamics of vaporized silicate melts for modeling volcanic outgassing and magma ocean atmospheres}

\correspondingauthor{Aaron Wolf}
\email{aswolf@umich.edu}
\author[0000-0003-2415-0508]{Aaron S. Wolf}
\affil{Earth and Environmental Sciences, University of Michigan, 1100 North University Avenue, Ann Arbor, MI 48109-1005, USA}
\author[0000-0002-2740-7965]{Noah J\"aggi}
\affil{Physics Institute, University of Bern, Sidlerstrasse 5, 3012 Bern, Switzerland}
\author[0000-0002-1462-1882]{Paolo A. Sossi}
\affil{Institute of Geochemistry and Petrology, Department of Earth Sciences, ETH Zurich, Clausiusstrasse 25, 8092 Zurich, Switzerland}
\author[0000-0002-0673-4860]{Dan J. Bower}
\affil{Center for Space and Habitability, University of Bern, Gesellschaftsstrasse 6, 3012 Bern, Switzerland}

\begin{abstract}
Silicate vapors play a key role in planetary evolution, especially dominating early stages of rocky planet formation through outgassed magma ocean atmospheres.
Our open-source thermodynamic modeling software ``VapoRock'' combines the MELTS liquid model \citep{Ghiorso1995} with gas-species properties from multiple thermochemistry tables \citep[e.g.][]{Chase1998}.
VapoRock calculates the partial pressures of 34 gaseous species in equilibrium with magmatic liquid in the system Si-Mg-Fe-Al-Ca-Na-K-Ti-Cr-O at desired temperatures and oxygen fugacities (\fO2, or partial pressure of \ch{O2}).
Comparison with experiments shows that pressures and melt-oxide activities (which vary over many orders of magnitude) are reproduced to within a factor of $\sim$3, consistent with measurement uncertainties. 
We also benchmark the model against a wide selection of igneous rock compositions including bulk silicate Earth, predicting elemental vapor abundances that are comparable (Na, Ca, \& Al) or more realistic (K, Si, Mg, Fe, \& Ti) than those of the closed-source MAGMA code (with maximum deviations by factors of 10-300 for K \& Si).
Vapor abundances depend critically on the activities of liquid components. The MELTS model underpinning VapoRock was calibrated and extensively tested on natural igneous liquids.
In contrast, MAGMA's liquid model assumes ideal mixtures of a limited set of chemically simplified pseudo-species, which only roughly approximates the non-ideal compositional interactions typical of many-component natural silicate melts.
Finally, we explore how relative abundances of \ch{SiO} and \ch{SiO2} provide a spectroscopically measurable proxy for oxygen fugacity in devolatilized exoplanetary atmospheres, potentially constraining \fO2 in outgassed exoplanetary mantles.
\end{abstract}

\keywords{Planet formation; Chemical thermodynamics; astrochemistry; planets \& satellites: atmospheres; planets \& satellites: composition}

% \section*{Potential Reviewers}
% \begin{itemize}
%     \item Laura Schaefer (lkschaef@stanford.edu), vapor modeling
%     \item Myriam Telus (mtelus@ucsc.edu), vaporization experiments
%     \item Conel Alexander (calexander@carnegiescience.edu), experience with MELTS and evaporation models
%     \item Ruslan Mendybaev (ramendyb@uchicago.edu) - thermodynamics of evaporation
%     \item Jacob B. Lowenstern (jlwnstrn@usgs.gov) volcanic outgassing on Earth + MELTS outgas modeling
%     \item emeritus Roger Hewins (hewins@scarletmail.rutgers.edu); volcanic outgassing with planetary applications
%     \item emeritus Lawrence Grossman (??); U chicago MELTS-based modeling of astro-vapors in disks
%     \item modeling spectra???
% \end{itemize}

% \section*{Major Tasks}
% \begin{itemize}
%     \item streamline code usage with cloud app
%     \item add fractional vaporization feature?
%     \item validate fractional vaporization on published labwork \citep[see][]{Schaefer2004}?
% \end{itemize}

% \listoftodos

% \section*{Standards}
% \begin{itemize}
%     \item 1 sentence per line please
%     \item use todo commands to leave notes for later work
%     \item Do not delete text wherever possible, use comments and eventually remove to graveyard file
    
% \end{itemize}

\section{Introduction}
%Silicate vapors receive little attention in many geoscience contexts because evidence for their participation in high temperature processes is often circumstantial. 
Silicate vapors play a key role in planetary evolution that spans the full range of %planetary
length- and time-scales, from the earliest proto-dust grains \citep[][]{hewins2005, fedkin2006vapor} to the catastrophic late-stage collisions \citep[][]{oneillpalme2008} and full-scale vaporization of rocky planets \citep[][]{Fegley1987, sossi2019evaporation}.\
%\subsection{Where does vapour occur?}
%here taken to mean those excluding the major volatile-bearing species comprising solely H, C, O, N and S
At near-equilibrium conditions, rocky silicate vapors---those incorporating variable oxidation states of the major and minor rock-forming elements (e.g. Si, Mg, Fe, Al, Ca, Na, K, Ti, Cr, and O)---are present above condensed liquid or solid phases, though usually at very low pressures at typical magmatic temperatures \citep[\(\leq10^{-10}\)~bar; e.g.][]{Sossi2018}.
In most situations, these vapors are thus insignificant when compared with the atmospheric pressures of major gaseous species, such as \ch{CO2} or \ch{H2O}, around rocky planets.

% when other external processes dominate, such as the biotically controlled surface conditions on Earth.

%\subsection{Where is evaporation relevant?}
However, silicate vapor is present in key high-temperature stages of planetary evolution.
Prior to the formation of the planets, the solar nebula---composed largely of \ch{H2}---condensed to form minerals and (in some cases) silicate liquids \citep[][]{Ebel2000a}.
These components subsequently accreted to form the terrestrial planets yielding compositional variability driven primarily by the range of condensation temperatures and strong radial \& vertical thermal gradients in the protoplanetary disk \citep[e.g.][]{wanke1988chemical, humayuncassen2000, sossi2022}. 
In later stages of planet formation, giant impacts were likely ubiquitous \citep[][]{morbidelli2012building} and readily produce global-scale melting and vaporization \citep[][]{Nakajima2015,Stewart2020}, reaching temperatures at which rock vapor atmospheres are likely to have formed \citep[e.g.,][]{charnoz2021, lock2018}.
As a result, rocky planets probably experienced at least one deep magma ocean stage, in which  exchange in both volatile and refractory elements between their mantle and silicate vapor atmospheres occurred \citep{Pahlevan2019,Stewart2020}.
During these magma ocean stages, silicate vapors coexist with abundant volatile gas species such as \ch{H2O} and \ch{CO2}, generating volatile-dominated atmospheres of several hundred bars \citep[e.g.][]{bower2022retention,gaillard2022redox}.
Finally, over the prolonged secular evolution of the planet, volcanic outgassing controls the long-term exchange of matter---primarily volatile and moderately volatile elements---between interiors and atmospheres \citep[][]{ortenzi2020,liggins2022}.

Further from home, some exoplanets with close-in orbits of their parent star, may undergo progressive vaporization and whole-scale erosion by atmospheric escape, claiming victims of both planetary migration and runaway stellar expansion in the red-giant phase \citep[][]{owen2019}.
Indeed, the atmospheres of hot rocky exoplanets, such as 55-Cancri-e, are prime targets for characterisation using the James Webb Space Telescope \citep[][]{zilinskas2022} and complementary ground-based techniques \citep{jindal2020}.
However, understanding how rock-forming elements are distributed between the vapor and condensed phase during these processes requires not only knowledge of the behaviour of the vapor phase, but, importantly, their chemical potentials (or activities) in silicate liquids.
Therefore, solving the puzzles of planetary evolution depends critically on our ability to model silicate vapors and their interplay with liquid and solid phases during condensation and vaporization processes.

% \awnote{@AARON: Check w/ Mark Ghiorso that the citation of ThermoEngine is correct}
Given the strong scientific motivation for better understanding silicate vaporization processes, we have developed and thoroughly validated a new code,  VapoRock \citep{VapoRock2021, Jaeggi2021}, which combines the MELTS liquid model---as implemented by the ThermoEngine%
\footnote{\href{https://gitlab.com/ENKI-portal/ThermoEngine}{https://gitlab.com/ENKI-portal/ThermoEngine}}
code \citep{ThermoEngine2022}---with gas-species properties from the thermochemistry tables of NIST-JANAF \citep[][]{Chase1998} and Lamoreaux and coworkers \citep[][]{Lamoreaux1984,Lamoreaux1987}, enabling coexisting vapor speciation calculations similar to those provided by the MAGMA code \citep[][]{Schaefer2004}.
VapoRock is capable of determining the abundances of gaseous species in equilibrium with a silicate magma, using the MELTS liquid model \cite[][]{Ghiorso1995}.
MELTS has been rigorously tested and validated by the petrology community over three decades and has also proven useful in simulating coexisting silicate vapors for magmatic planetary outgassing \cite[in the private code of][]{Ito2015}, chondrule evaporation \citep{fedkin2006vapor, alexander2002}, and condensation of the early solar nebula \cite[in the no longer operable VAPORS code,][]{Ebel2000b}.
Currently, the closed-source tool MAGMA is the primary code available to the community that is capable of predicting partial pressutes of gas species in equilibrium with a silicate magma ocean.
Other, mostly closed-source codes have also been used in the literature or reported at conference talks, though with significantly less extensive experimental and theoretical validation work supporting their use \citep{Hin2017,Shornikov2019,Schlichting2022, vanBuchem2022}.
%\awnote{I am just going to ref LavAtmos as an olive branch to their Arxiv}
It is important to note that one capability available in MAGMA (but not currently offered by VapoRock) is fractional vaporization, which simulates progressive vaporization and removal, as may be relevant for heavily irradiated planets, and also provide an additional potential avenue for experimental validation \citep[e.g.,][]{Schaefer2004}.
Fortunately, VapoRock is under active development and usage so this feature will be explored in detail in a subsequent study, and new features can always be found at the open public repository for the project.%
\footnote{\href{https://gitlab.com/ENKI-portal/vaporock}{https://gitlab.com/ENKI-portal/vaporock}}

In this work, we establish the foundation for vaporization modeling with VapoRock by exhaustively comparing the predictions of partial pressures and activities using VapoRock and MAGMA (both the 2004 and 2022 versions) with those determined in experimental studies.
Accordingly, we demonstrate significantly improved results for VapoRock in line with experimental constraints, owing to the greater accuracy of its underlying liquid model for natural silicate melts.
Finally, we use the model to explore the utility of the observed SiO to \ch{SiO2} ratio as a simple proxy for oxygen fugacity in devolatilized exoplanet atmospheres.

%Thus a major benefit of our project will be the extension of the open-source ThermoEngine code to treat magma-vapor systems as a new tool for the astronomy and planetary science communities.

% <!-- # [[202101111354]] Liquid/Vapor equilibrium thermochemical database -->
%Modeling coupled liquid-vapor evolution relies upon an accurate thermochemical database that spans the compositional range of important geologic and planetary systems.
%We adopt the chemical system of MELTS \citep{Ghiorso1995}, a petrologic modeling tool which has stood the test of time and has also proven useful in simulating condensation of the early solar nebula as part of the no longer operable VAPORS code \citep[][]{Ebel2000b}.

% <!-- # [[202101121012]] Thermochemical tables for vapor species modeling -->

\section{Method}

\subsection{Modeling vaporized silicates overview}
\label{sec:database_overview}
%<!-- # [[202101050640]] Modeling vaporized silicates overview -->
To quantitatively study rock vaporization processes, we require a complete set of accurate thermodynamic models for condensed and vapor phases sampling broad ranges in composition and temperature.
It is critical that these solution models faithfully represent geologically realistic compositions, including  the dominant refractory rock-forming elements (e.g. Si, Mg, Fe, etc.) as well as the minor moderately volatile elements (e.g. K, Na), which play key roles in vaporization due to their strong preference for the vapor phase.
Fortunately, this burden is eased by the typically low-pressure conditions of vaporization, where thermodynamic models of the gas phase can neglect the complications introduced by compression---with the notable exception of super-critical vapors in high-energy giant impacts \cite[e.g.][]{lock2018}.
As such, we rely on the ideal gas approximation to extend to arbitrarily low partial pressures and mixed compositions.
Calculating the composition, speciation, and abundance of the gas phase likewise requires an extensive and detailed thermodynamic database of vapor species, conveniently provided by standard thermochemical tables.
In this paper, we will focus exclusively on the vapor-limited ``outgassing'' regime of vaporization, where molten rock dominates the system coexisting with an exceedingly small amount of equilibrium vapor.
%, saving exploration of the vapor-dominated regime relevant to condensation from a stellar nebula for a forthcoming publication.
A system is vapor-limited when the amount of vapor (by mass) is vastly outweighed by condensed phase(s) (in our case the silicate liquid), ensuring that the overall bulk composition is nearly equal to that of the liquid, with vapor negligibly affecting the elemental budget of the system.
%equilibrium state of the system is largely set by the condensed phase(s) alone, allowing us to model these systems in two steps, first considering only condensed phase(s) and then adding vapor species as a second-order correction.
%With this approach, we can use existing thermodynamic models, like the equilibration algorithm embedded in MELTS, to establish condensed-phase equilibrium, after which we can then determine the coexisting abundances of vapor species.
%The task is even further simplified at higher temperatures where the condensed phase is fully molten.
%In this case, condensed-phase equilibrium is trivial, since the composition and abundance of liquid are equal to the bulk composition and total mass of the system, respectively.
The conditions of equilibrium are therefore controlled by the liquid, which dictates the chemical potentials of the system ($\vec{\mu}$) and thus the corresponding vapor abundances, which are calculated in a separate secondary step.
Due to their multi-component nature, silicate liquids can span a wide range of oxygen fugacities in nature, from extremely reducing (in equilibrium with metallic iron and silicon) to highly oxidized (with all iron in the ferric state).
While many vaporization codes (including MAGMA) %impose additional constraints on the cation-oxygen ratio of the vapor 
use the vaporization reactions themselves to predict \fO2, %we know from both theory and experiment that the full range of naturally occurring values are possible for any silicate liquid.
%To remain entirely general, 
VapoRock %thus
retains \fO2 as a user input (independent variable), reflecting the external equilibrium conditions of the liquid being simulated.
One might similarly expect pressure to be a model input, but it is actually fully predicted by the vapor-liquid equilibrium, as a sum over the partial pressures of each vapor species.
Thus the necessary inputs for a VapoRock calculation are the liquid (system) composition, the temperature, and the \fO2, from which vapor species abundances can be predicted assuming vapor-liquid equilibrium.

%Vapor species abundances (i.e., partial pressures) are finally determined directly from these chemical potentials, providing a complete picture of all the stable phases in the system.

\subsection{Modeling silicate liquid/vapor equilibrium}
\label{sec:liq_vapor_equil}
Our thermochemical model relies on the silicate liquid MELTS model \citep{Ghiorso1995}, and thus describes a subset of the \ch{SiO2}--\ch{MgO}--FeO--\ch{Fe2O3}--\ch{Al2O3}--CaO--\ch{Na2O}--\ch{K2O}--\ch{TiO2}--\ch{Cr2O3} system, to which liquid and vapor compounds are readily converted using stoichiometric constraints.
%  <!-- [[202101121125]] -->
% <!-- strip out tbl refs since not working yet in open lab-book (FIX LATER) -->
The MELTS model itself adopts liquid endmember components (see Table \ref{tab:database}) that were chosen to maximize model accuracy 
%(during both calibration and extrapolation)
while retaining a simple quadratic mathematical form (appropriate to regular solution theory).
To couple with MELTS, we adopt a description of the potential vapor species in this chemical system, considering all lone elements and energetically favorable molecular species.
This chosen set of system components and vapor species is comparable to the volatile-free model space of the MAGMA code \citep{Schaefer2004, Fegley1987}.
%which additionally includes H and C for volatile-containing magmas
%The data describing these vapor species is obtained from a variety of (overlapping) thermochemical databases, namely NIST-JANAF \citep[][]{Chase1998} and that of Lamoreaux and coworkers \citep[][]{Lamoreaux1984,Lamoreaux1987}. The agreement between them is within ~1 \% relative in Gibbs Free Energies at 1500 K, which is sufficiently good agreement for these modeling purposes.

\begin{deluxetable}{@{}llc@{}}[!htbp]
\renewcommand{\arraystretch}{1.3}
\caption{Thermochemical database of VapoRock \label{tab:database}
}
\tablehead{
\colhead{system components (c)} & \colhead{liquid components (l)} & \colhead{vapor species (v)}\\
}
\startdata
\ch{SiO2} & \ch{SiO2} & Si, \ch{Si2}, \ch{Si3}, SiO, \ch{SiO2}\\
\ch{MgO} & \ch{Mg2SiO4} & Mg, \ch{Mg2}, MgO\\
\ch{FeO}, \ch{Fe2O3} & \ch{Fe2SiO4}, \ch{Fe2O3} & Fe, FeO\\
\ch{Al2O3} & \ch{Al2O3} & Al, \ch{Al2}, \ch{Al2O}, AlO, \ch{Al2O2}, \ch{AlO2}\\
\ch{CaO} & \ch{CaSiO3} & Ca, \ch{Ca2}, CaO\\
\ch{Na2O} & \ch{Na2SiO3} & Na, \ch{Na2}, NaO\\
\ch{K2O} & \ch{KAlSiO4} & K, \ch{K2}, KO\\
\ch{TiO2} & \ch{TiO2} & Ti, TiO, \ch{TiO2}\\
\ch{Cr2O3} & \ch{MgCr2O4} & Cr, CrO, \ch{CrO2}, \ch{CrO3}\\
- & - & O, \ch{O2}\\
\enddata
\tablecomments{Cr- and Ti-based species are exclusive to JANAF implementation, as they are not included in \cite{Lamoreaux1984,Lamoreaux1987}. Fe thermochemical data from JANAF is used in Lamoreaux calculations due to the critical importance of Fe to most natural igneous systems. Outside this exception, vapor databases are not mixed to allow independent comparison of model predictions. 
} 
\end{deluxetable}

%\subsection{Thermochemical tables for vapor species modeling}
%\label{thermo_vapo}

High-temperature thermochemical models rest upon comprehensive databases of vapor species energetics.
Given low density conditions where the ideal gas approximation holds, a complete thermodynamic system model can be built solely from Gibbs energy expressions for each gas species.
These models are typically presented as tabular data evaluated on a grid of temperatures, suitable for interpolation lookup, or as empirical equations as a function of temperature.
The most widely used example is the NIST-JANAF thermochemical database \citep[][]{Chase1998}, which contains an exhaustive set of materials described by both lookup tables and analytic expressions (using the Shomate equations; see Appendix \ref{app:shomate-eqns}).%
\footnote{IVTAN thermochemical tables \citep[][]{gurvich1989} cover a large set of compounds (derived from Russian-based research efforts), but full digital access is rather challenging.}
A more limited dataset focused specifically on metal- and metal oxide gas species is provided by \cite{Lamoreaux1984} and \cite{Lamoreaux1987}, which uses a polynomial description of relative Gibbs energy changes ($\Delta G / RT$).
As one might hope, the energetics described by each of these independent data sources are in reasonable agreement (typically at the level of $\pm100 J$), sufficient for vaporization modeling purposes.
In VapoRock, the option to toggle between both the JANAF- and Lamoreaux-based expressions for gas species thermodynamic properties is provided, so as to prevent mixing of data sources and to explore any potential predictive differences that may arise.
%This agreement is particularly important as most of these data sources are incomplete in some respect, and thus some mixing of sources is typically required to obtain a full description of all compounds of interest.

%\subsection{Chemical potentials for liquid-dominated system w/ MELTS}
% <!-- [[202101130617]] Chemical potentials for liquid-dominated system w/ MELTS -->
%In liquid-dominated systems---where the liquid phase houses the majority of every element---the conditions of equilibrium are entirely dictated by the energetics of the liquid phase.
%In typical applications, state of the system in terms of bulk composition, pressure, temperature, and oxygen fugacity are externally imposed or evolved through time, and thus the properties of the liquid state set the chemical potential of the system.
Silicate liquid properties are given by the 10-component (volatile-free) MELTS liquid model \citep{Ghiorso1995}, which rapidly calculates the chemical potentials of the system as a function of current system conditions (i.e. temperature, molar composition, and $f{\ch{O2}}$).
Since the thermodynamic properties of the liquid are not sensitive to ideal gas-relevant pressure levels (P $\lesssim$ 1 kbar), the calculation of liquid chemical potentials is performed assuming a nominal pressure of 1 bar, yielding an approximation that remains accurate as long as the ideal gas approximation itself holds.%
\footnote{Due to weak pressure sensitivity, updating the nominal liquid pressure by setting it equal to the total gas species pressure results in a negligible correction factor.}
This computation provides the chemical potentials in terms of the liquid endmember components, but can be readily converted to oxide system components by relying on the linearity of chemical potentials (see Appendix \ref{app:stoic} for details).

% One might similarly expect pressure to be a model input, but it is actually fully predicted by the vapor-liquid equilibrium.
% Given a known or imposed fO2, the chemical potentials of the liquid fully determine the parital pressure of each vapor species and the total pressure is merely the sum over all species.
% Since the thermodynamic properties of the liquid are not sensitive to ideal-gas-relevant pressure levels (P << 1 kbar), the calculation is performed a nominal 1 bar pressure with the knowledge that this approximation is accurate as long as the ideal gas approximation itself holds.
% Put another way, updating the nominal liquid pressure to equal the total gas species pressure results in an entirely negligible and unnecessary  correction to the species abundances.
% There are certainly some super-critical ultra-high P/T systems for which the ideal gas approximation breaks down (like in the moon-forming impact) but they represent a minority of potential applications and would require a distinct modeling approach.
% Thus the necessary inputs for a VapoRock calculation are the liquid (system) composition, the temperature, and the fO2, from which vapor species abundances can be predicted assuming vapor-liquid equilibrium.

\subsection{Calculating equilibrium vapor abundances}
% <!-- # [[202006110721]] Equilibrium Vapor Species Abundances above Condensed Phs -->
% <!-- Equilibrium abundances of every vapor species above a condensed phase are readily calculated given a model of each pure species at 1 bar. -->
In the vapor-limited regime, thermodynamics provides a straightforward framework for predicting vapor compositions, where the abundance of each vapor species is independently adjusted to equalize its chemical potential with the liquid.
Using mass-balance constraints, we write the balanced chemical reaction that forms the $i^{th}$ gaseous species from the liquid by exchange of basic system components, e.g., the oxides with additional oxygen consumed or produced as needed:

\begin{equation}
    \phi_{iv} = \sum_{j} \nu_{ij} c_j + \nu_{i\ch{O2}} c_{\ch{O2}}
    \label{eq:chemical_formula}
\end{equation}
where $\phi_{iv}$ is the chemical formula of the $i^{th}$ vapor species, $c_j$ and $c_{\ch{O2}}$ are the vector of basic system components (oxides plus oxygen), whereas $\nu_{ij}$ and $\nu_{i\ch{O2}}$ give the stoichiometry of the vapor species in question, expressing its composition in terms of system components.
With the law of mass action, the corresponding equilibrium condition is written:
\begin{equation}
    \begin{split}
        \mu_{iv} &= \sum_{j} \nu_{ij} \mu_j + \nu_{i\ch{O2}} \mu_{\ch{O2}}\\
        \mu^0_{iv} + RT\ln P_{iv} &= \sum_{j} \nu_{ij} \mu_j + \nu_{i\ch{O2}} \cdot ( \mu_{\ch{O2}}^0 + RT\ln f_{\ch{O2}})
    \end{split}
    \label{eq:chemical_potential}
\end{equation}
where $P_{iv}$ is the equilibrium partial pressure of vapor species $i$ expressed in bars, $f$\ch{O2} is the oxygen fugacity, and $\mu_{iv}$, $\mu_j$, and $\mu_{\ch{O2}}$ are the chemical potentials of vapor species $i$, component $j$, and molecular oxygen, respectively.
Critically, the pure-phase energetics of each gas species, expressed in terms of their molar Gibbs energy or chemical potential $\mu^0_{iv}$ \& $\mu_{\ch{O2}}^0$, are given by the 1 bar thermochemical databases (JANAF and Lamoreaux) detailed in Sec. \ref{sec:database_overview}.
This expression of equilibrium imposes equality of chemical potentials for gas species $i$ in terms of system components that are freely exchanged with the coexisting liquid.
The above formulation uses chemical potentials directly, but is equivalent to the activity-focused formulation of \cite{Sossi2018}.
Rearranging, we obtain the governing equation for the abundances of each vapor species in terms of its partial pressure:
\begin{equation}
    \log P_{iv} = \frac{1}{RT \ln{10}} \left[ \sum_{j} \nu_{ij} \mu_j + \nu_{i \ch{O2}}\mu_{\ch{O2}}^0 - \mu_{iv}^0 \right] + \nu_{i\ch{O2}} \log f_{\ch{O2}}
    \label{eq:partial_pressure}
\end{equation}
where the factor of $\ln10$ is included for convenience to convert from natural logarithms to more intuitive base-10 logarithms for the pressure and oxygen fugacity.
% <!-- {#eq:vapor-abundances} -->
The equilibrium abundance of each vapor species is thus determined by environmental conditions (temperature and oxygen fugacity), the liquid composition (which dictates the chemical potentials of the system), and the known 1~bar thermal properties of each vapor species ($\mu_{iv}^0$ and $\mu^0_{\ch{O2}}$).

Calculation of species abundances using the governing expression above, rests upon the simplified behavior of ideal gases in the low density limit, where species in a mixed vapor do not interact with one another.
%The ideal gas approximation imposes a simple link between pressure and temperature effects through the ideal gas law, enabling experimental constraints at a 1~bar reference pressure to trivially extend to arbitrarily low-density pressure conditions.
%For this application, the primary benefit of the ideal gas approximation is the absence of 
This enables us to model mixed vapors without needing to consider compositional mixing terms, ensuring that the fugacity of any gaseous species is simply equal to its partial pressure, $f_{iv}=P_{iv}$, enabling the abundance of each vapor species to be determined independently of all others.
Additionally, the ideal gas law ($P_{iv} = n_{iv} R T$) relates partial pressures to molar number densities for each species, $n_{iv}$, allowing us to trivially determine the bulk elemental composition of the vapor.

\section{Results}

Before discussing potential applications of the VapoRock code, it is critical to understand the accuracy of the partial pressures it computes, and compare them to independent experimental constraints. 
This may be achieved in two primary ways, i) direct comparison of measured partial vapor pressures and ii) calculation and assessment of activity coefficients in the silicate melt.
These comparisons can be applied to both direct experimental measurements as well as benchmark calculations made by VapoRock and the MAGMA model for a representative set of natural magmatic compositions. 
For these benchmarks, we use two separate versions of the MAGMA code: MAGMA 04---the 2004 version that originally introduced K-vaporization \citep{Schaefer2004}; and MAGMA 22---the updated 2022 version designed to improve the accuracy of K and Na outgassing (as presented in \citealt{Martin2017}; \citealt{Jiang_etal2019}).
By comparing to both models, we explore the general capabilities of the MAGMA modeling framework as well as the specifics of each calibration.

\subsection{Thermodynamic analysis of experimental measurements}

Up to now, our thermodynamics presentation has been focused on direct assessment of the chemical potentials (or partial molar Gibbs energies) of the liquid/vapor system.
For many experimental vaporization studies, it is convenient to instead view vaporization from an alternate thermodynamic activity model perspective.
These two approaches are equivalent and are easily converted from one to the other using the thermodynamic definition of activity:
\begin{equation}
    \mu_i = \mu_i^0 + RT \ln{a_i}
    \label{eq:activity_chempot}
\end{equation}
By comparing this expression with that of an ideal solution, the activity ($a_i$) acts as an effective mole fraction of the $i^{\rm th}$ species or component. %simplifying the calculation of thermodynamic properties for complex non-ideal materials like silicate melts.
The factor that relates activity to mole fraction is the activity coefficient, $\gamma_i$ (where $a_{i} = \gamma_{i} \, x_{i}$), which expresses the degree of non-ideality through how much it deviates from unity.
The activity coefficient can formally adopt any non-negative value ($0 \le \gamma_{i} < +\infty$), but typically maintains values less than but of order unity ($0.1 \lesssim \gamma_{i} \lesssim 1$), with important exceptions for silicate liquids (see discussion in Sec. \ref{sec:liquid_model}). 
The activity coefficient can be determined by relating the partial pressure of the $i^\textrm{th}$ gaseous species in the complex system, \textit{p}$_{i}$ (e.g. as predicted by VapoRock or determined in experiments),  to the expected partial pressure in an idealized system, $p_i^0$:
\begin{equation}
    \gamma_{i} = p_{i}/p_{i}^{0}
    \label{eq:p_pnought}
\end{equation}
This idealized system contains a liquid with the same composition (i.e., mole fraction) as the true mixed liquid, but that behaves as an ideal solution %with an activity that is just equal to the mole-fraction for the oxide component of interest
(i.e., $a_i^0 = x_i$).
The idealized pressure of the species, $p_i^0$, can then be directly determined from thermochemical tables by rearranging the law of mass action, which relates the equilibrium constant for a particular reaction to a ratio of the activities of products and reactants (see Appendix \ref{app:mass-action-law}). 
The ratio of the true pressure to this idealized pressure then yields the activity coefficient of the component in the liquid that is in equilibrium with the gaseous species of interest, for instance providing an estimate of the $\gamma$\ch{SiO2(\textit{l})} based on the known \textit{p}\ch{SiO(\textit{g})}.

These vapor-constrained liquid activity coefficients are readily examined in terms of their temperature-dependence, allowing quantitative comparison between theoretically modeled values and experimentally measured ones.
Integration of the Clausius-Clapeyron equation (or its equivalent the Van 't Hoff equation for activities) over the temperature range, \textit{T}, under a set of simplifying assumptions appropriate to liquids and gases yields an equation of the form:
\begin{equation}
    \ln\gamma_{i} = A/T + B
    \label{eq:lnP_ATB}
\end{equation}
where A and B are constants related to the enthalpy and entropy of the reaction, respectively. 
The simplifying assumptions are merely that the heat capacity of the liquid and gas are temperature-independent, $\Delta$C$_P$ = 0, producing constant values for the enthalpy and entropy of formation for gaseous species from the liquid.
The values of A and B are determined by plotting ln($p_{i}/p_{i}^{0}$) (equivalent to ln$\gamma_{i}$ from Eq.~\ref{eq:p_pnought}) as a function of reciprocal temperature. 
Thus, we are able to both model the temperature evolution of vapor abundances as well as compare theory and experiment in terms of their differences in the implied enthalpy and entropy of the vaporization reaction.

Validation of the vapor species abundances directly is achieved by contrasting our results with Knudsen Effusion Mass Spectrometry (KEMS) measurements of the partial pressures for vapor species composed of rock-forming elements.
The goal in our comparison is to characterize the log partial pressure difference between the VapoRock modeled and experimentally measured vapor pressures.
% \begin{equation}
%     \delta p_{i} = \log(p_{i}^\textrm{VR}/p_i^\textrm{exp})
%     \label{eq:pi_exp_VR}
% \end{equation}
% with small values reflecting good agreement between VapoRock (VR) modeled and experimentally (exp) measured vapor pressures.
It should be noted that the derivation of equilibrium partial pressures using the KEMS method, although well-suited for simple solids \citep[e.g.][]{sergeev2019, copland2001, costa2017,drowart2005}, becomes complicated for multi-component silicate melts due to their tendency to react with and creep out of the Knudsen cell \cite{Bischof2021}.
Moreover, the fragmentation of molecular gas species pollute the mass spectrum, rendering identification of the parent molecule ambiguous.
Alternatively, validation can be performed by comparing the inferred activity coefficients for the silicate liquid, which directly control the vapor species abundances.
For activity-based comparisons, we can assess the log activity coefficient difference between the modeled VapoRock and experimentally determined values.
% \begin{equation}
%     \delta \gamma_i =  \log(\gamma_{i}^\textrm{VR}/\gamma_{i}^\textrm{exp})
%     \label{eq:gamma_exp_VR}
% \end{equation}
% with smaller values indicating better agreement between modeled and measured activity coefficients.

As discussed in section \ref{sec:liq_vapor_equil}, the tabulated values of the free energies of gas species from different databases are typically within $\pm$ 100 J of one another.
This observation, combined with the approximation that the gas behaves ideally, means that much of the discrepancy between experimental data and models is expected to arise from differences in the activity models for the liquid in most situations (see discussion in Sec.~\ref{sec:liquid_model}).

\subsection{Experimental validation in complex synthetic systems}
\label{sec:complex-synth-valid}

%Direct laboratory constraints on silicate melt (or glass) vaporization are fairly sparse in the literature due to the inherent experimental challenges.
%In more complex systems with four (or more) components 
The most extensive body of literature for complex synthetic silicate liquids is on silicate `slags', which are important for industrial and mining applications.
%Slag glass vaporization experiments thus provide a potential source for experimental benchmarking of silicate vaporization modeling.
The inherent challenge in comparing with experimental measurements of vapors above slags is that most industry-relevant melting and vaporization experiments examine compositions that do not readily typify terrestrial magmas, and thus do not represent relevant tests of the model for astronomical and geological applications.
%The inherent challenge associated with a comparison between vapors above slags and those calculated by VapoRock is that most industry-relevant melting and vaporization experiments examine compositions that do not readily typify terrestrial magmas, and thus do not represent relevant tests of the model in the areas of interest.
%are focused on rather simplified or idealized compositions rather than natural samples.
%This is a particular problem since the MELTS liquid model is %specifically
%built on natural igneous rock compositions, and loses considerable accuracy when comparing with compositions that lie very far from the calibration data. 

\begin{figure}[ht]
    \centering
    \includegraphics[width=0.49\textwidth]{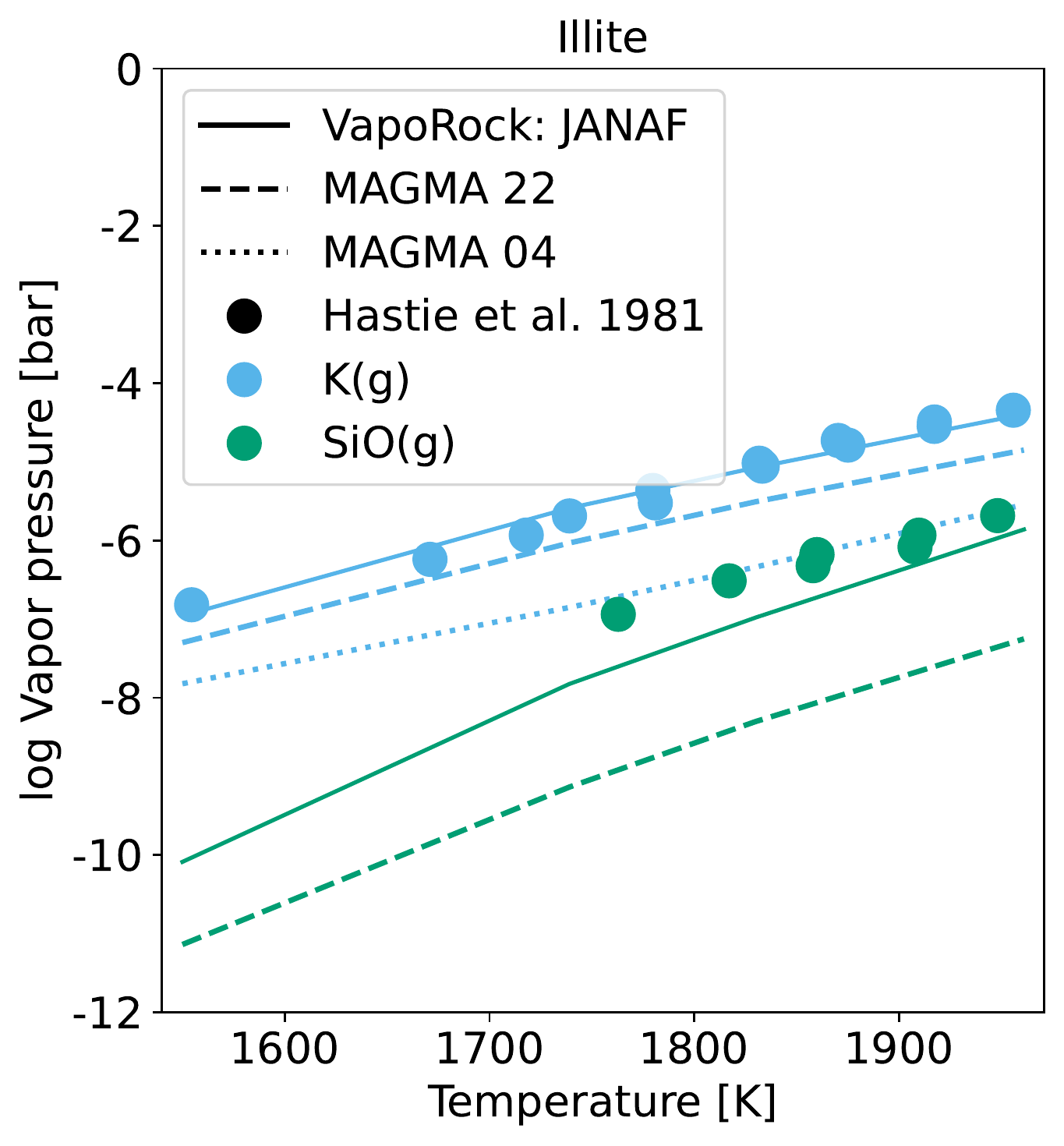}
    \includegraphics[width=0.49\textwidth]{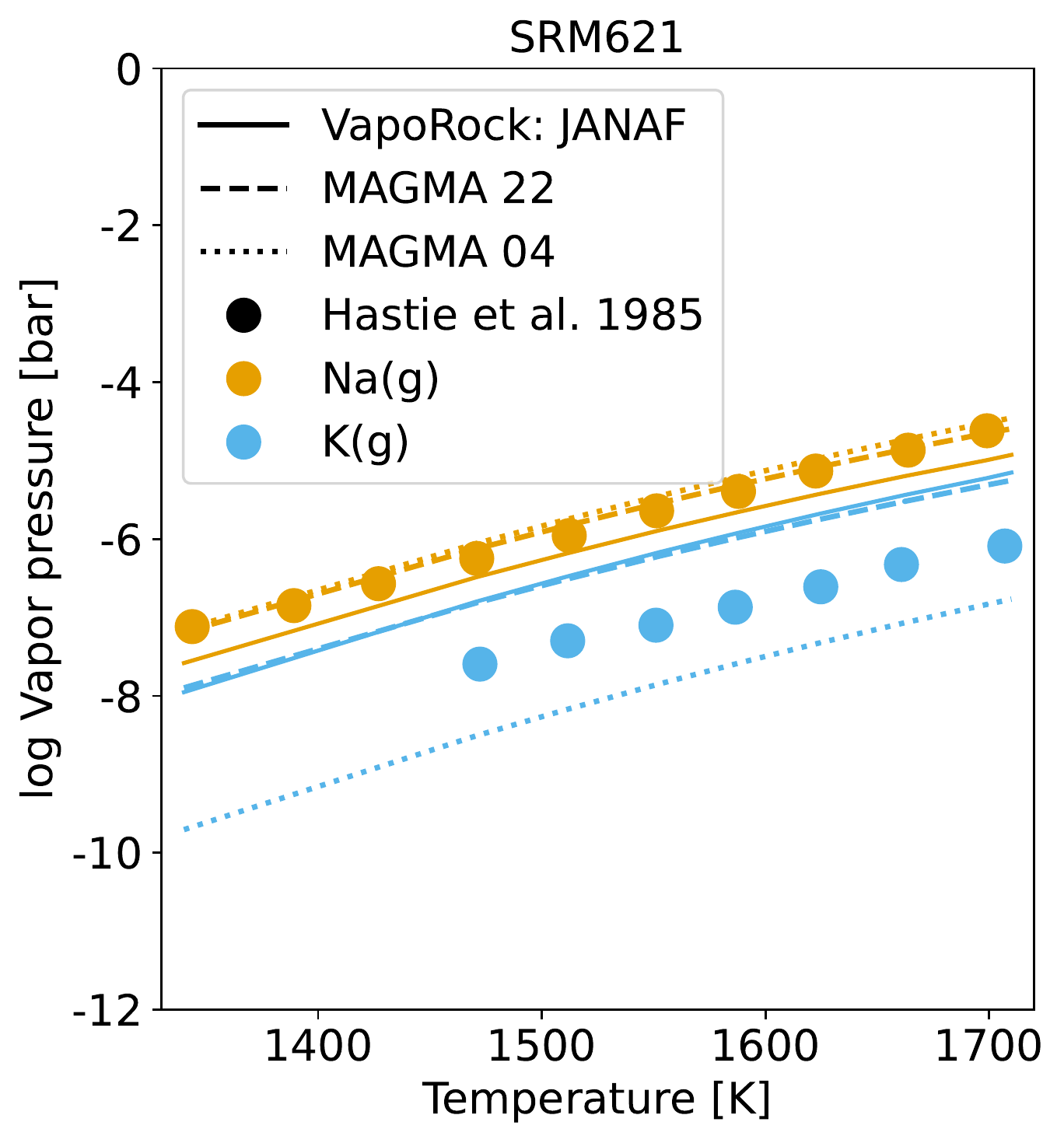}
    \caption{VapoRock comparison for vaporization experiments on synthetic illite and SRM 621 glasses. Experimental determinations from \cite{Hastie1981} and \cite{Hastie1985} are shown as color-coded points, and VapoRock predictions are shown in solid lines (using the JANAF gas species database), while MAGMA predictions are shown in dashed and dotted lines \citep[from the current 2022 version and][respectively]{Schaefer2004}. Glass compositions in wt\% are illite:~(\ch{SiO2}~=~60.2, \ch{Al2O3}~=~26.0, MgO~=~2.1, \ch{Fe2O3}=~~4.4, \ch{K2O}~=~7.4) and SRM~621:~(\ch{SiO2}~=~71.39, \ch{Al2O3}~=~2.78, MgO~=~0.27, \ch{Fe2O3}~=~0.04, \ch{K2O}~=~2.02, \ch{Na2O}~=~12.75, \ch{CaO}~=~10.75). VapoRock results based on Lamoreaux thermochemical data coincide with JANAF results and are not shown \citep{Lamoreaux1984,Lamoreaux1987}. Uncertainties on KEMS-derived absolute partial pressures are within 50 \% in chemically inert systems \citep{drowart2005} but likely higher for silicate melts.
    \label{fig:Hastie-validation}}
\end{figure}

We therefore limit our direct comparison with experiments to the most geologically-relevant slag glasses carried out in a series of studies by J. W. Hastie and colleagues in the 1980s, %on mixed alkali glasses, These experiments studied a range of synthetic glass compositions, the most complex and geologically-relevant
namely, the NIST standard glass SRM 621 and an illite-based glass (see Fig.\ref{fig:Hastie-validation} caption for compositions).
Partial pressures of SiO(\textit{g}), Na(\textit{g}), \ch{O2}(\textit{g}), $\&$  K(\textit{g}) above SRM 621 liquid and illite liquid were simultaneously measured by the KEMS method over temperature ranges from 1340~-~1710~K and 1550~-~1960~K, respectively \citep{Hastie1981, Hastie1985, Hastie1986}. 
Based on the recorded oxygen fugacities%
\footnote{It should be noted that MAGMA is capable of predicting \fO2 values that nicely match these experiments by relying on stoichiometric constraints not currently implemented in VapoRock, which instead uses measured \fO2 as an input.}
and the known glass composition, we use VapoRock to predict the partial pressures of SiO, Na, and K as shown in Fig.\ref{fig:Hastie-validation}.
    
%As depicted in Figure \ref{fig:Hastie-validation}, the points represent measurements reported in \citep{Hastie1985}, while the lines show model predictions from VapoRock.
%For this calculation, the experimental oxygen fugacity trend was , enabling continuous modeling of these data across the temperature range.
%The dashed lines represent the results using the reported glass composition for the sample, showing the correct relative volatility of the Na and K vapor species, but under-predicting the volatility contrast.
%For these high-volatility species, the vapor abundance predictions follow the same general trend and are generally within 1/2 to 1 order of magnitude of their experimental values.
%For comparison, we also show in dotted lines the predicted volatility curves for a modified sample composition, where the $\ch{Na_2O}$ and $K_2O$ sample compositions were adjusted (keeping all other oxides fixed) to best fit the observed vapor abundances 
%This procedure provides an alternative way of conceptualizing the differences between these experiment and model results, indicating that the volatility differences correspond to a necessary adjustment of the Na and K2O abundances by a factor of 1.57 and 0.15, respectively.

For the illite composition, the level of agreement in partial pressures between VapoRock and KEMS is excellent, with \textit{p}K being reproduced to within 50 \% across the entire temperature range (Fig.~\ref{fig:Hastie-validation}). 
The agreement with \textit{p}SiO is also good, particularly at the highest temperatures at which its measurement is more precise, with VapoRock predicting values within a factor 2, while MAGMA 22 underpredicts \textit{p}SiO by more than an order of magnitude. 
This concordance for VapoRock slightly worsens at lower temperatures to a factor 5, likely owing to poor counting statistics on SiO during the KEMS measurements. 
This translates into a near-constant $\gamma$\ch{SiO2} of 0.80 $\pm$ 0.01 for VapoRock, while the data of \citet{Hastie1981} imply decreasing $\gamma$\ch{SiO2} from 3.7 at 1550 K to 1.5 at 1960 K. 
For $\gamma$KO$_{0.5}$, both the KEMS data and the VapoRock output can be fit to the equation ln($\gamma$KO$_{0.5}$)~=~-29380/\textit{T} + 4.13, reflecting their excellent level of correspondence.
%\awnote{@PS, @NJ: Check activity values here to be sure they are still right given updates to SRM composition.}\njnote{only SRM, will forward data to PS} \psnote{Thanks guys - updated!}

Vapor pressures of \textit{p}Na and \textit{p}K measured above SRM 621 glass, in comparison, show a similar offset to the values predicted by VapoRock. 
Na abundances are underpredicted by VapoRock by a factor of 3 - 4, while K is overpredicted by a factor of 2 - 3. 
Activity coefficients for NaO$_{0.5}$ are given by ln($\gamma$NaO$_{0.5}$)~=~-18520/\textit{T} + 1.85 (VapoRock) and ln($\gamma$NaO$_{0.5}$)~=~-18320/\textit{T} + 2.40 \citep{Hastie1981}. Similarly, we find ln($\gamma$KO$_{0.5}$)~=~-26690/\textit{T} + 5.64 (VapoRock) and ln($\gamma$KO$_{0.5}$)~=~-24080/\textit{T} + 1.95 \citep{Hastie1981}.
Overall, this level of agreement is actually a net improvement over the MAGMA model \citep{Schaefer2004}, which shows somewhat better values for Na (generally matching the experiments) but performs substantially worse for K in MAGMA 04, where abundances are under-predicted by 1 to 1.5 orders of magnitude for both illite and SRM~621 glass compositions (Fig~\ref{fig:Hastie-validation}, Table~\ref{tab:MAGMA_comparison}).
This situation is largely imporved with the updated K model of MAGMA 22, which matches VapoRock for SRM 621 but still remains slightly low by a factor of $\sim2$ for illite.

\subsection{Experimental validation in natural systems}
\label{sec:lunar-valid}
The 1970s saw a major scientific push to study lunar samples returned from the Apollo missions, spurring a host of experimental studies aimed at understanding the geological history of the lunar surface and its interior.
In particular, the observation that lunar mare basalts are highly depleted in moderately-volatile elements (notably Na and K) relative to terrestrial rocks was interpreted to indicate a high-temperature past involving significant vaporization, either during their eruption \citep{ohara1970} or on a whole-Moon scale \citep{ringwood1970}.\\

\begin{figure}[ht]
    \centering
    \includegraphics[width=0.49\textwidth]{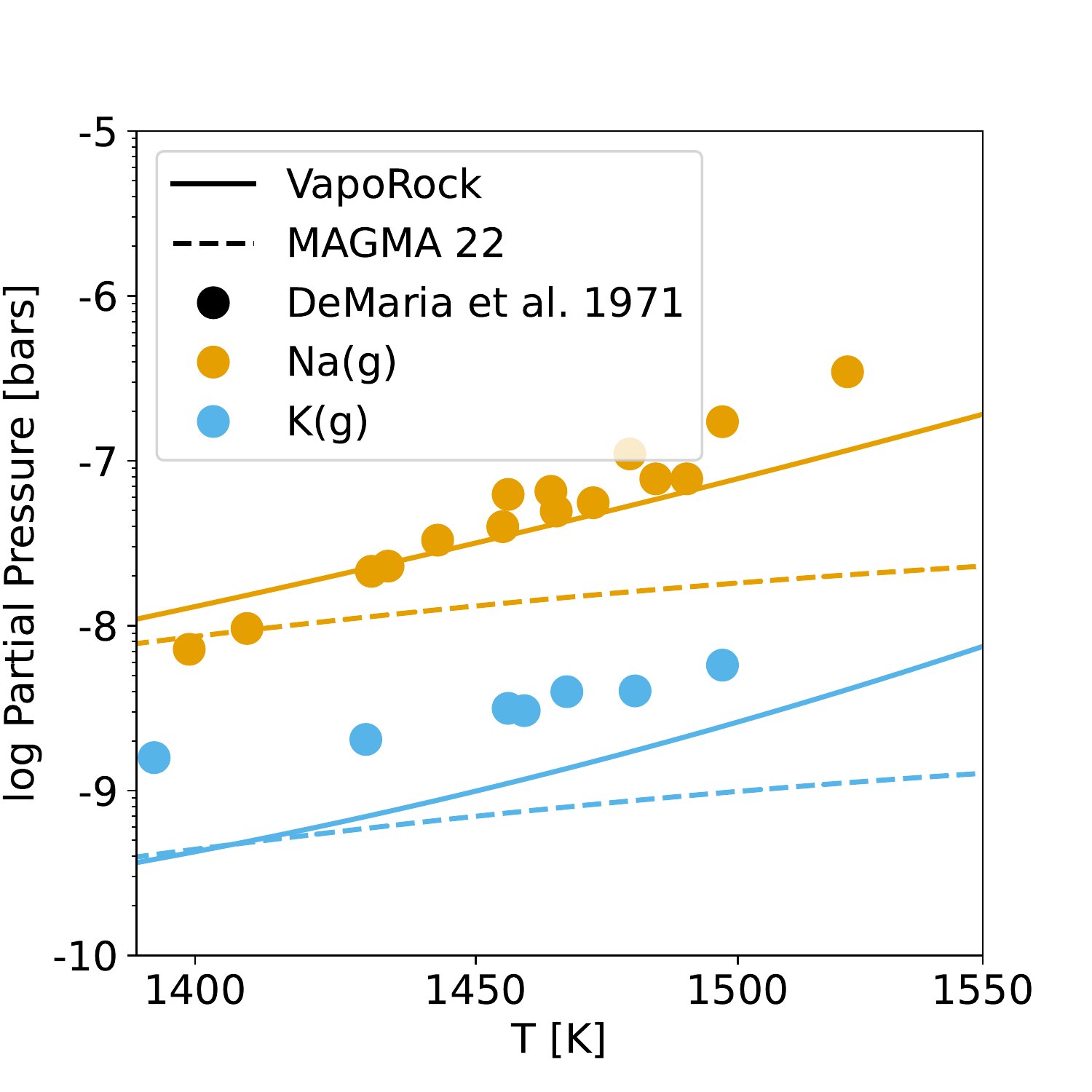}
    \caption{Vapor pressures of Na(\textit{g}) (green) and K(\textit{g}) (orange) during evaporation of lunar basalt (Apollo sample 12065, see composition in Table \ref{tab:MAGMA_comps}) as determined by KEMS \citep[][points]{deMaria1971} and as calculated by VapoRock (full lines) and MAGMA (2022 Version available at the time of writing, dashed lines).}%\njnote{ changed back to temperature scale for Reviewer Nr.1 - old invT version is still in figures folder}}
    \label{fig:DeMaria-validation}
\end{figure}

As reviewed in \cite{Sossi2018}, vaporization experiments were carried out on lunar samples by \cite{demaria1969}, \cite{deMaria1971}, \cite{Naughton1971}, \cite{Gooding1976}, and \cite{Markova1986}.
Of these, by far the most systematic study is that of \cite{deMaria1971}, in which two Low-Ti mare basalts, 12022 and 12065, were heated in Re Knudsen cells to 2270~K and 2500~K, respectively (see Table~\ref{tab:MAGMA_comps} for the composition of sample 12065).
The partial pressures of Na, K, \ch{O2}, Fe, SiO, Mg, Mn, Cr, Ca, Al, AlO, TiO and \ch{TiO2} were determined as a function of temperature by mass spectrometry.
As the \textit{p}\ch{O2} values are required as an input for VapoRock, we limit our comparison to Na and K only (see Fig.~\ref{fig:DeMaria-validation}).
This is because these were the two elements present in measurable quantities in the vapor phase over the temperature range \citep[1396~-~1499~K;][]{deMaria1971} at which \textit{p}\ch{O2} was determined. \\

As shown in Fig.\ref{fig:DeMaria-validation}, the partial pressures of Na and K given by VapoRock are in broad agreement with those determined by \cite{deMaria1971}. 
%However, in detail, there are differences. 
%At 1400 K, VapoRock overpredicts \textit{p}Na by a factor of 2, while at 1500 K it underpredicts \textit{p}Na by the same factor.
On average, VapoRock nicely captures the observed Na(g) partial pressures, though it somewhat underestimates the temperature dependence, with a maximum factor of two deviation from measured \textit{p}Na at the temperature extremes of 1400 K and 1500 K.
%\awnote{@PS: Why does this analysis pretend that the temp extremes are 1400 and 1500K when data spans from 1300 to 1550K? Need to fix or explain for this whole paragraph.} \psnote{Yep good point. I think I extrapolated the O2 measurements to lower temperatures than actually measured (1396 K) in order to use VapoRock to model the results. But I think we can just remove these points, as the data don't fit well in any case. @Noah, could you replot Fig. 2 just down to the 1396 K temperature point?}
Additionally, the observed rate of increase of \textit{p}K with temperature is well-reproduced by VapoRock, though the absolute values are roughly three times lower than observations, improving to within a factor two by 1500 K.
Expressions derived for the melt activity coefficients of the cation-normalized oxide components in VapoRock are ln($\gamma$NaO$_{0.5}$)~=~-20264/\textit{T} + 2.37 and ln($\gamma$KO$_{0.5}$)~=~-28743/\textit{T} + 3.36.
The ratio $\gamma$KO$_{0.5}$/$\gamma$NaO$_{0.5}$ in lunar basaltic melt predicted by VapoRock therefore varies from 0.006 (1400 K) to 0.009 (1500 K). 
At the highest temperatures (1500 K), where measurements of partial pressures are most reliable, the KEMS-derived $\gamma$KO$_{0.5}$/$\gamma$NaO$_{0.5}$ is also 0.01. 
As KEMS-derived partial pressures are frequently not known to better than $\sim$50 \% relative \citep{deMaria1971, drowart2005} a conservative error on this figure would be log($\gamma$KO$_{0.5}$/$\gamma$NaO$_{0.5}$) = -2 $\pm$ 0.3, which agrees with VapoRock predictions over the entire temperature range.
We also plot the predictions of MAGMA 22, finding poorer agreement with the data, especially at higher temperatures.
For both Na and K, the predicted temperature slope using MAGMA 22 is much lower than VapoRock, causing the Na values to agree only at 1400K with deviations growing to factors of 10 to 20 at 1500 K.
For K, it shows a consistent mismatch with the data by a factor of $\sim$5 across the full temperature range.
We can thus see that for experimentally measured natural lunar basalts, VapoRock does a good job of capturing alkali vaporization behavior, especially at more geologically-relevant higher temperatures where analytical uncertainties are also smaller.

%These experimental efforts had both the aim of constraining the partial pressures of gas species in the equilibrium vapor phase of lunar basalts and other terrestrial or meteorite samples, as well as understanding their chemical trajectory as they are modified by fractional vaporization.
When interpreting these comparisons with direct experimental constraints, the challenges in obtaining quantitatively useful measurements of partial pressures from multi-component liquids are not to be underestimated.
Differential fractional vaporization of the most volatile components of the silicate liquid cause the composition of the liquid to evolve as evaporation progresses.
These changes are nearly unavoidable since collection times needed for precise measurements are on par with evaporation timescales, and thus measurements are unavoidably following a moving target in terms of both vapor abundances and liquid composition.
This is compounded by uncertainties in ionization cross sections, fragmentation of complex molecules, and contribution of cell material to the mass spectrum of the analysed gas species in KEMS \citep{drowart2005}.
In light of these challenges, the level of agreement to within a factor of 2-3 in calculated partial pressures by VapoRock for lunar basaltic samples is on par with the uncertainties in the measurements themselves, building confidence in the model's predictive power for natural magmatic systems.

\subsection{Model benchmark against existing MAGMA code}

The MAGMA code \citep{Fegley1987,Schaefer2004} is the 
%present benchmark 
community's standard tool for the prediction of partial pressures of metal- and metal oxide components above silicate melts, and has been extensively employed to understand the evaporation of the Earth \citep{Schaefer2009}, the Moon \citep{canup2015}, and super-Earths \citep{Schaefer2009}.
It employs the `equilibrium constant' approach, in which partial pressures are computed according to equilibrium constants at a given temperature and pressure.
The liquid activity model differs from that employed by VapoRock (MELTS) in that a number of intermediate melt components are devised that are then assumed to mix ideally in order to simulate non-ideal mixing.
The resulting liquid and gas phase proportions are given by a combination of mass action and mass balance to yield unique solutions for the speciation of the vapor and of the equilibrium melt.

To assess the predictive differences between VapoRock and MAGMA, we examine four of the complex natural rock compositions simulated by the MAGMA code \citep[see][their Table 5]{Schaefer2004} and two additional compositions (Bishop Tuff and Type B CAI), covering a broad cross-section of igneous rocks (Table \ref{tab:MAGMA_comps}).

\begin{deluxetable}{@{\extracolsep{4pt}}lrrrrrrrr@{}}[!htbp]
\label{tab:MAGMA_comps}
\tablewidth{0pt} 
\tablecaption{Model compositions in wt\% for a wide range of igneous rocks.}
\tablehead{
\colhead{} & \colhead{} & \colhead{Alkali} & \colhead{} & \colhead{} & \colhead{Bishop} & \colhead{Type B} & \colhead{Lunar} & \colhead{} \\% \multicolumn{2}{c}{Lunar glass}\\
\colhead{Oxide} & \colhead{Tholeiite} & \colhead{basalt} & \colhead{Komatiite} & \colhead{Dunite} & \colhead{Tuff} & \colhead{CAI} & \colhead{basalt} & \colhead{BSE} %\colhead{green} & \colhead{orange}
}
\startdata                                              
\ch{SiO2}   & 50.71 & 44.80 & 47.10 & 40.20 & 75.60  & 44.78  & 46.87   & 45.97 \\ %& 47.1  & 45.16 & 38.18 \\  
\ch{MgO}    & 4.68  & 11.07 & 29.60 & 43.20 & 0.25   & 17.05  & 7.82    & 36.66 \\ %& 9.2   & 17.38 & 14.15 \\  
\ch{Al2O3}  & 14.48 & 13.86 & 4.04  & 0.80  & 13.00  & 11.41  & 10.05   & 4.77  \\ %& 9.8   & 7.81  & 6.1 \\   
\ch{TiO2}   & 1.70  & 1.96  & 0.24  & 0.20  & 0.21   & 0.00   & 3.34    & 0.18  \\ %& 3.1   & 0.53  & 9.78 \\  
\ch{Fe2O3}  & 4.89  & 2.91  & 0.00  & 1.90  & 0.00   & 0.00   & 0.00    & 0.00  \\ %& 0.0   & 0.00  & 0.00 \\  
\ch{FeO}    & 9.07  & 9.63  & 11.5  & 11.9  & 1.10   & 0.00   & 19.76   & 8.24  \\ %& 20.0  & 19.98 & 22.93 \\  
\ch{CaO}    & 8.83  & 10.16 & 5.44  & 0.80  & 0.95   & 26.76  & 10.73   & 3.78  \\ %& 10.1  & 8.45  & 7.59 \\  
\ch{Na2O}   & 3.16  & 3.19  & 0.46  & 0.30  & 3.35   & 0.00   & 0.27    & 0.35  \\ %& 0.24  & 0.14. & 0.36 \\  
\ch{K2O}    & 0.77  & 1.09  & 0.09  & 0.10  & 5.55   & 0.00   & 0.073   & 0.04  \\ %& 0.063 & 0.05  & 0.17 \\
\ch{Cr2O3}  & 0.00  & 0.00  & 0.00  & 0.00  & 0.00   & 0.00   & 0.00    & 0.00  \\ %& 0.5   & 0.52  & 0.69\\
\multicolumn{8}{c}{Oxides not considered in VapoRock}\\\hline
\ch{MnO}    & 0.22  & 0.17  & 0.22  & 0.20  & 0.00   & 0.00.  & 0.256   & 0.14  \\ %& 0.26    & 0.00   & 0.00 \\  
\ch{H2O}    & 1.04  & 0.73  & 0.00  & 0.40  & 0.00   & 0.00   & 0.00    & 0.11  \\ %& 0.00    & 0.00   & 0.00 \\  
\ch{P2O5}   & 0.36  & 0.55  & 0.05  & 0.10  & 0.00   & 0.00   & 0.00    & 0.02  \\ %& 0.00    & 0.00   & 0.00 \\  
\ch{NiO}    & 0.00  & 0.00  & 0.00  & 0.00  & 0.00   & 0.00   & 0.00    & 0.24  \\ %& 0.00    line %& 0.00   & 0.00 \\ 
\hline
Total       & 99.91 & 100.12& 100.04& 100.10& 100.01 & 100.00 & 100.36  & 100.50  \\\hline %& 100.02 & 99.95\\  \hline
Refs.        & \multicolumn{1}{c}{[1]}   & \multicolumn{1}{c}{[2]}   & \multicolumn{1}{c}{[3]}   & \multicolumn{1}{c}{[4]}   & \multicolumn{1}{c}{[5]}   & \multicolumn{1}{c}{[6]}   & \multicolumn{1}{c}{[7]} & \multicolumn{1}{c}{[8]}% & [9] & [10]   & [10]
\enddata
\tablecomments{
[1] \cite{Murase1973}; [2] \cite{Carmichael1974}, p. 501 \#1; [3] Arndt (2002, personal communication to B. Fegley); [4] \cite{Macdonald1972}); [5] \cite{Gualda2012}; [6] \cite{Richter2007}; [7] Lunar basalt, Apollo sample 12065 \cite{Maxwell1971,Meyer2004}, \cite{deMaria1971}; [8] Bulk Silicate Earth (BSE) \cite{oneill1998}. % [9] Lunar basalt, Apollo sample 12065 \cite{Smales1971}[10] Means of compositions of \cite{delano1986}.
}
\end{deluxetable}

\begin{figure}[ht]
    \centering
    \includegraphics[width=1\textwidth]{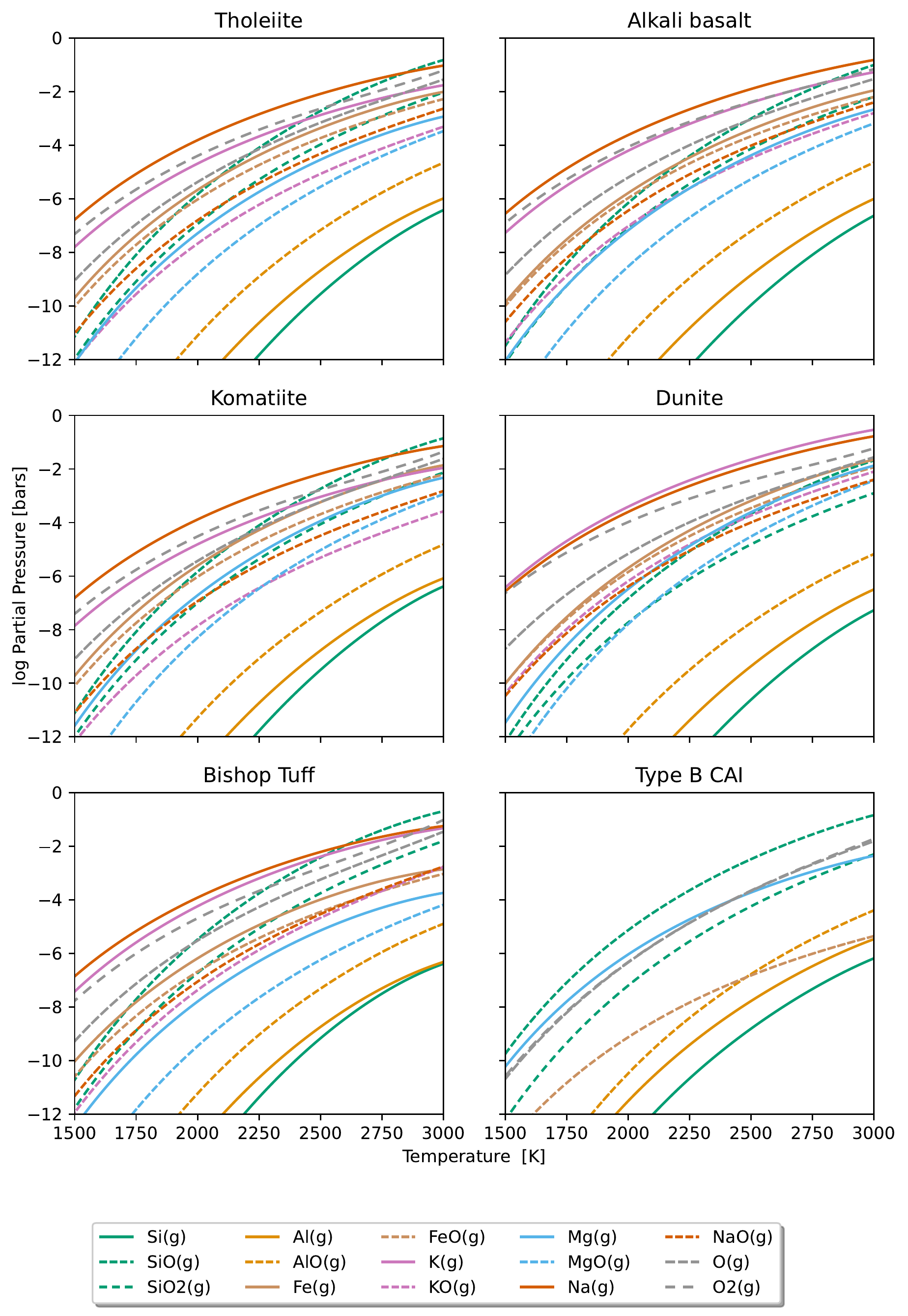}
    \caption{Vapor pressures for dominant gas species calculated using VapoRock for multiple igneous rock compositions from \cite{Schaefer2004}, Bishop Tuff composition from \cite{Gualda2012}, and Type B CAI from \cite{Richter2007}. Partial pressures computed at temperature-dependent oxygen fugacities derived from  f\ch{O2} of the same compositions. Gas species are color-coded by primary cation type (see legend) and line styles reflect the number of oxygen atoms for each species, with more broken lines indicating more oxygens.}
    \label{fig:Pfrac-Schaefer}
\end{figure}

Resulting partial pressures of the 15 most abundant gas species are shown in Fig. \ref{fig:Pfrac-Schaefer}, computed over a temperature range from 1500 K to 3000 K, where we impose the predicted \fO2 value from MAGMA 22 to ensure a fair comparison.%
\footnote{The oxygen fugacity is reported relative to the iron-wüstite buffer ($\Delta$IW) in terms of log$_{10}$ unit deviations in \fO2, where log\fO2 (IW) = -28776.8/\textit{T} + 14.057 + 0.055$\times$(\textit{P}-1)/\textit{T} - 0.8853$\times$ln(\textit{T}), with \textit{P} in bar and \textit{T} in K \citep{oneillpownceby1993, hirschmann2008}.} 
% \awnote{Removed O2 buffer footnote here. can be moved elsewhere if needed} \psnote{I think this footnote is very important so that others can see how we calculate it, so I put it back :)}
%(at a constant relative $\log f$\ch{O2} of $\Delta$IW=+1.5)
The results indicate that Na and K are typically the most abundant gas species over most compositions up to $\sim$2750~K, before SiO(\textit{g}) becomes more abundant. 
This cross-over occurs at log(\textit{p}$_i$) near -2, but in detail it varies depending on the bulk composition of the liquid.
In general, the temperature of the cross-over is negatively correlated with the \ch{SiO2} content of the melt, with the Bishop Tuff (\ch{SiO2} = 75.6 wt. \%; 3.35 wt. \% Na$_2$O) reaching the transition at 2500 K, whereas for Dunite (\ch{SiO2} = 40.2 wt. \%; 0.30 wt. \% Na$_2$O) it is in excess of 3000 K.
However, it should be noted that the cross-over temperature is also positively correlated with alkali content of the liquid.
Because SiO$_2$ and the alkalis typically increase in tandem during igneous differentiation as broadly incompatible elements, this conspires to keep the cross-over temperature relatively constant for a wide range of terrestrial igneous rocks.
Higher \fO2 would result in a crossover occurring at higher temperatures as Na becomes even more volatile with respect to Si.
The next most abundant species are typically \ch{O2}, O, Fe, and Mg. 
The most refractory elements are Al, Ti, and Ca and are only present in the vapor as minor species at levels below 10$^{-6}$ bar, even at 3000 K.\\

To facilitate comparison of the predicted vapor compositions, we convert the vapor species abundances to mole fractions of each element present in the gas phase using the ideal gas law at constant temperature (1900 K) and \fO2 for each liquid composition (Table \ref{tab:MAGMA_comparison}).
Elemental abundances are presented relative to Na, the dominant vapor species, and the ratio of predicted values from VapoRock over those from MAGMA are given for each composition.
Due to modest systematic differences in predicted Na abundances between the two models, we also present the Ca-normalized ratios, which benefit from more regular and comparable behavior in both models and reduce disagreement for most elements.
A clear mismatch of $\sim$2-3 orders of magnitude is observed between the partial pressures of total K (relative to either Na or Ca) between VapoRock and the original MAGMA code (hereafter MAGMA 04).  % $\Sigma$K/$\Sigma$Ca
%which are of the order of 10$^{-4}$ to 10$^{-3}$ in the MAGMA code, but closer to 10$^{-1}$ to 10$^{0}$ in VapoRock.
This significant underestimation of the vapor pressures of K species in MAGMA 04 is a widely-known issue (compare \textit{p}Na/\textit{p}K in the Moon composition of \cite{canup2015} to that of \cite{visscher2013}), and was first pointed out by \citep{Ito2015}.
In response, the liquid activity and vapor equilibrium constant models used to predict K-bearing gas species partial pressures were updated in \citet{Martin2017} and \citet{Jiang_etal2019}, respectively.
We subsequently employ this newest version of the MAGMA code (as of October, 2022) to calculate vapor pressures, reported in Table \ref{tab:MAGMA_comparison} as MAGMA 22.
The resultant differences in K partial pressures between VapoRock and MAGMA 22 are vastly reduced; and typically agree to within a factor of 5.
Using the Ca-normalized ratios, we can see that VapoRock predicts lower Na values by a factor of only 3-5 compared to MAGMA, a level of agreement that is acceptable for moderately volatile elements given the overall match to external experimental constraints (see Sec.~\ref{sec:complex-synth-valid}-\ref{sec:lunar-valid}).

\begin{deluxetable*}{llDDDDDDDD}[!htbp]
\label{tab:MAGMA_comparison}
\tablecaption{Comparison of predicted elemental abundances for vaporized lavas}
\tablehead{
\multicolumn{2}{c}{Composition} &     \multicolumn{2}{c}{Na}  &     \multicolumn{2}{c}{K}  &     \multicolumn{2}{c}{Al} 
    & \multicolumn{2}{c}{Si}  & \multicolumn{2}{c}{Ca} 
    & \multicolumn{2}{c}{Ti} 
    & \multicolumn{2}{c}{Fe} & \multicolumn{2}{c}{Mg}\\
\colhead{} & \colhead{}         & \multicolumn{2}{c}{$10^0$} & \multicolumn{2}{c}{$10^{-4}$} & \multicolumn{2}{c}{$10^{-9}$}
    & \multicolumn{2}{c}{$10^{-3}$}   &  \multicolumn{2}{c}{$10^{-8}$}   
    & \multicolumn{2}{c}{$10^{-6}$} 
    &  \multicolumn{2}{c}{$10^{-3}$} &  \multicolumn{2}{c}{$10^{-4}$}
}
\decimals
\startdata
    %%%%%%%%%%%%%%
    \hline \hline
    \bf Tholeiite
        & VapoRock           & 1   & 1262 & 13.43 & 4.43 & 1.31 & 19.24 & 13.83 & 1.76 \\
    \bf (IW+2.9)   
        & MAGMA 04           & 1   &    2.11 &  8.77 & 2.35 & 0.83 & 1.68 & 4.10  & 0.45 \\
        & MAGMA 22           & 1   &    613  & 9.70  & 0.83 & 0.99 & 1.79 & 4.62  & 0.52 \\ 
    \noalign{\smallskip}
    \hline             \; \; \it [M/Na]
        &\; \; \emph{ratio} (04)& 1  & 598 & 1.5 & 1.9 & 1.6 & 11  & 3.4 & 3.9 \\ 
        &\; \; \emph{ratio} (22)& 1  & 2.1 & 1.4 & 5.3 & 1.3 & 11  & 3.0 & 3.4 \\
    \noalign{\smallskip} \; \; \it [M/Ca]
        &\; \; \emph{ratio} (04)& .6 & 379 & 1.0 & 1.2 & 1.0 & 7.3 & 2.1 & 2.5 \\
        &\; \; \emph{ratio} (22)& .8 & 1.6 & 1.0 & 4.0 & 1.0 & 8.1 & 2.3 & 2.6 \\ 
    %%%%%%%%%%%%%%
    \hline \hline
    \bf Alkali Basalt
        & VapoRock           & 1   & 2470 & 7.52 & 1.37 & 0.96 & 11.76 & 7.59 & 1.45 \\
    \bf (IW+3.2)
        & MAGMA 04           & 1   &    1.88 & 2.68  & 0.35 & 0.42 & 0.67 &  1.26 &  0.37 \\
        & MAGMA 22           & 1   &  1043   & 3.44 & 0.15 & 0.61 & 0.81 & 1.67 & 0.53 \\ 
    \noalign{\smallskip}
    \hline             \; \; \it [M/Na]
        &\; \; \emph{ratio} (04)& 1 & 1314 & 2.8 & 3.9 & 2.3 & 18 & 6.0 & 3.9 \\ 
        &\; \; \emph{ratio} (22)& 1 & 2.4 & 2.2 & 9.1 & 1.6 & 15 & 4.5 & 2.7 \\
    \noalign{\smallskip} \; \; \it [M/Ca]
        &\; \; \emph{ratio} (04)&  .4 & 574.8 & 1.2 & 1.7 & 1.0 & 7.7 & 2.6 & 1.7 \\
        &\; \; \emph{ratio} (22)& .6 & 1.5 & 1.4 & 5.8 & 1.0 & 9.2 & 2.9 & 1.7 \\
    %%%%%%%%%%%%%%
    \hline \hline   
    \bf Komatiite
        & VapoRock           & 1   & 1134 & 10.44 & 5.16 & 0.88 & 3.03 & 17.33 & 7.62 \\
    \bf (IW+2.8)
        & MAGMA 04           & 1   &    1.87 & 5.79  &  1.27 & 1.03 & 0.21 &  4.53 & 3.70 \\
        & MAGMA 22         & 1   &    610 & 7.86 & 0.52 & 1.47 & 0.27 & 6.26 & 5.73 \\ 
    \noalign{\smallskip}
    \hline             \; \; \it [M/Na]
        &\; \; \emph{ratio} (04)& 1 & 606 & 1.8 & 4.1 & 0.9 & 14 & 3.8 & 2.1  \\ 
        &\; \; \emph{ratio} (22)& 1 & 1.9 & 1.3 & 9.9 & 0.6 & 11 & 2.8 & 1.3 \\ 
    \noalign{\smallskip} \; \; \it [M/Ca]
        &\; \; \emph{ratio} (04)&  1.2 & 709.8 & 2.1 & 4.8 & 1.0 & 17 & 4.5 & 2.4  \\
        &\; \; \emph{ratio} (22)& 1.7 & 3.1 & 2.2 & 17 & 1.0 & 19 & 4.6 & 2.2 \\
    %%%%%%%%%%%%%%
    \hline \hline
    \bf Dunite
        & VapoRock           & 1   & 15306 & 2.01 & 0.28 & 0.13 & 1.33 & 8.70 & 6.73 \\
    \bf (IW+3.4)
        & MAGMA 04           & 1   &   29.20 & 0.40 & 0.08 & 0.08 &  0.03 &  1.26 & 1.45 \\ 
        & MAGMA 22         & 1   &  1859 & 0.47 & 0.03 & 0.12 & 0.04 & 1.57 & 2.02 \\ 
    \noalign{\smallskip}
    \hline             \; \; \it [M/Na]
        &\; \; \emph{ratio} (04)& 1 & 524 & 5.0 & 3.5 & 1.6 & 44.3 & 6.9 & 4.6 \\ 
        &\; \; \emph{ratio} (22)& 1 & 8.2 & 4.3 & 9.3 & 1.1 & 33 & 5.5 & 3.3 \\ 
    \noalign{\smallskip} \; \; \it [M/Ca]
        &\; \; \emph{ratio} (04)&  .6 & 323 & 3.1 & 2.2 & 1.0 & 27.3 & 4.2 & 2.9 \\
        &\; \; \emph{ratio} (22)& .9 & 7.6 & 3.9 & 8.6 & 1.0 & 31 & 5.1 & 3.1 \\
    %%%%%%%%%%%%%%
    \hline \hline
    \bf Bishop Tuff
        & VapoRock      & 1 & 4354 & 12.97 & 12.60 & 0.21 & 26.62 & 5.15 & 0.72 \\
    \bf (IW+2.6)
        &  MAGMA 22     & 1 & 1042 & 3.44 & 0.15 & 0.61 & 0.81 & 1.67 & 0.53 \\
    \noalign{\smallskip}
    \hline             \; \; \it [M/Na]
        &\; \; \emph{ratio} (22)& 1 & 4.2 & 3.8 & 84 & 0.3 & 33 & 3.1 & 1.4 \\
    \noalign{\smallskip} \; \; \it [M/Ca]
        &\; \; \emph{ratio} (22)& 2.9 & 12 & 11 & 244 & 1.0 & 95 & 9.0 & 3.9 \\
    \noalign{\smallskip}
    \hline \hline
\enddata
\tablecomments{
Predictions from VapoRock are compared with MAGMA \citep[][here as MAGMA 04]{Schaefer2004} and the latest currently available code (MAGMA 22) calculated at 1900 K for various compositions (assuming the same $\Delta$IW for VapoRock as returned by MAGMA 22) and are reported relative to Na (with scale factors for each element in column header). The ratio of elemental abundances in VapoRock relative to MAGMA (both versions) are reported for each composition using Na-scaled values, {\it [M/Na]}, as well as preferred Ca-scaled values, {\it [M/Ca]}, which limit bias from systematic offsets in Na abundances between the two models.}
\end{deluxetable*}

The agreement between MAGMA and VapoRock for the abundances of refractory and major elements ($\Sigma$M/$\Sigma$Ca where M = Al, Si, Ti, Fe, or Mg) is considerably better than K or Na, but important differences still exist with modeled abundance ratios ranging from 0.5 to 20 (and as high as 300 for Si using the most recent version of MAGMA). 
%Thus, the VapoRock code invariably predicts higher $\Sigma$M/$\Sigma$Na, likely indicating that MELTS-derived activities for NaO$_{0.5}$ are systematically lower than those determined by the MAGMA code.
%Re-normalising the results to Ca reveals that 
Both MAGMA and VapoRock yield similar $\Sigma$Al/$\Sigma$Ca values for the vapor, while those for Ti, Si, Fe, and Mg show larger discrepancies in ratios of roughly 3 to 5 (up to 30 for Ti and 300 for Si). 
As stated in \cite{Schaefer2004}, the construction of the MAGMA code solution model does not allow for activity coefficients that are greater than unity. 
For most major elements, this is not a critical setback, as the melt oxide components containing Ca, Mg, Si, and Al all have activity coefficients $<$1 (cf. supplement Table 1).
%\ref{tab:simplesystems}).
However, this is not the case for Fe and Ti, whose melt oxide activity coefficients, $\gamma$FeO and $\gamma$\ch{TiO2} are typically of the order of 1~-~3 and 3~-~5, respectively, in basaltic melts at 1673 K \citep[cf.][]{oneilleggins2002, borisov2020}.
As such, one would expect the partial pressures of Ti- and Fe-bearing species to be underpredicted to the same extent in the MAGMA code (see Eq.~\ref{eq:p_pnought}), similar to what is observed (Table \ref{tab:MAGMA_comparison}).
The discrepancy of up to a factor of 30 for Ti remains, despite the normalisation to Ca.
%This leaves Si as the sole element discrepant between the two codes without a clear explanation.
%However, as evident from \cite[][their Fig. 6]{Schaefer2004}, 
Furthermore, there is evidence that the MAGMA code underpredicts $\gamma$\ch{SiO2} by a factor of 3 for CAI-type liquid compositions based on the experimental measurements of \citet{Richter2007} \citep[see][their Fig. 6]{Schaefer2004}, a problem that persists and even worsens in the MAGMA 22 code, where Si abundances are typically off by and order of magnitude, but can be as high as $\sim$300 for silica-rich Bishop Tuff.
%Thus, while we cannot definitively prove the cause for this divergence, 
%Thus, previously published studies also see accuracy limitations to the predictions of the MAGMA code for refractory rock-forming elements that are consistent with the offsets 
The calculated differences between VapoRock and MAGMA for vaporization of natural silicate liquids thus persists across multiple versions of the MAGMA model and are similarly observed in previous theoretical and experimental studies.
%presented in the test compositions in Table \ref{tab:MAGMA_comparison}.

\section{Discussion}
\subsection{Key differences in the underlying liquid model}
\label{sec:liquid_model}
To better understand the typically observed three- to ten- (up to 300) fold differences between the predictions of VapoRock and MAGMA for the vapor pressures of certain elements, especially for silica-related quantities, it is useful to contrast the form of the underlying MAGMA liquid model with that of the VapoRock (MELTS) model.
The MAGMA code relies upon the Ideal Mixing of Complex Components (IMCC) liquid model, which postulates the existence of a large number of representative species in the silicate liquid which mix together ideally \citep{Fegley1987}.
In effect, the liquid is itself mathematically represented as an ideal gas composed of liquid species each possessing their own fixed enthalpies and entropies of formation, which is quite convenient as it treats both the liquid and vapor phases symmetrically.
There is a cost to this simplicity, however, as it only roughly approximates both the temperature-dependence and the fundamental interactions between liquid components in complex many-component silicate liquids.

%\awnote{Moved section here to appendix, and placed a reference in the following paragraph. I think this section is now 2 paragraphs (1 pg!) shorter and flows much better.}

Within the IMCC modeling framework underlying the MAGMA code, pseudo-species act as a set of discrete samples of the Gibbs energy surface in composition space, deflecting its entropy and enthalpy to match experimental constraints at those locations.
It further relies on the sensible form of the ideal entropy of mixing to act as a smooth interpolation function between samples.
%This method is actually quite similar to the approximate thermodynamic computational approach of the PerpleX thermodynamic code \citep[ConnolyXXX], which relies on a grid of pseudo-compounds to sample and discretely represent the Gibbs energy surface of mineral solution phases.
%One of the key limitations of this approach, however, is that it relies on a dense grid of samples to accurately capture the non-ideal behavior of the solution, which is dominated by complex multi-component interactions for silicate melts (see Appendix \ref{app:mixed-silicate-thermo} for details).
%In the case of the silicate liquid model underlying MAGMA, it is fundamentally impractical to obtain the experimental constraints needed to model the pseudo-species required to adequately sample the high dimensional composition space.
%The IMCC model at the heart of the  and presupposes ideal interactions between these pseudo-species.
%This limitation is clearly demonstrated by the fact that the MAGMA code is entirely calibrated on data from liquid binary and ternary subsystems, without any species representing higher-order mixtures.
One of the major limitations of the IMCC model is thus its reliance on pseudo-species to fully capture the non-ideal shape of the Gibbs energy surface, which is only constrained where pseudo-species happen to lie in composition space.
Since pseudo-species within MAGMA only sample bounding binary and ternary subsystems, they fundamentally cannot capture non-ideal interactions for mixed compositions representing most natural liquids (for a detailed discussion of the multi-component interactions that dominate silicate liquid properties, see Appendix \ref{app:mixed-silicate-thermo}).
%(e.g. ranging from primitive basalts to highly evolved rhyolites for typical planetary silicate melts).
For the large 10+ dimensional compositional space used by MELTS, the IMCC framework would need a dense sampling of pseudo-species to accurately capture the non-ideal shape of the liquid's Gibbs energy surface.
Overcoming this limitation is unfortunately neither pragmatic nor even theoretically possible for silicate liquids, owing to the fact that the requisite number of sample points grows exponentially with the number of components in the liquid.%
%Obtaining a sufficiently dense coverage of the compositional space is far beyond our current experimental capabilities due to the vast region that must be covered 
\footnote{Assuming modest average sample spacing of 1/3 (or 1/4) mole fraction, the requisite number of pseudo-species is 220 (715) for 10 components. To capture many-component interactions, we need at least one pseudo-species in the interior of the 10-component space. This sampling-scheme grow exponentially, requiring 92,378 samples for a 10-component liquid. The actual MAGMA model from 2004 has only 46 species, falling short by 1 to 4 orders of magnitude.}
%But beyond the impracticality of obtaining sufficient experimental constraints, the required mineral compounds with the necessary intermediate compositions don't even exist in nature, since mineral structures are limited in their set of allowable cation substitutions.

% It is for this very reason that the MELTS model relies on coexisting mineral phases to constrain the compositional derivatives of the energy surface (rather than the direct values) to be able to directly model natural silicate liquids.

%the necessary minerals used to derive these constraints simply do not exist in nature.
%The model coefficients for each pseudo-species are provided by the well-characterized melting properties of mineral compounds, where the enthalpy, entropy, fusion temperature, and heat of fusion are known for the solid.
%But the unique thing about silicate liquids is the `omni-component' nature of the solution, where it easily accommodates arbitrary amounts of all 10+ oxides in any proportion (at high temperatures relevant to vaporizaton).
%We have no minerals that posses such broad solution properties; they are instead bound to compositional subsystems.
%Try as we may, the IMCC model simply cannot provide the requisite coverage of composition space required to yield accurate descriptions of non-ideal mixing in complex natural silicate liquids.

The IMCC approach thus relies on pseudospecies to fully sample the mixing properties of the liquid and is only able to accurately represent the non-ideal mixing properties for small deviations away from the binary and ternary subsystems where it is trained.
While this seriously impacts its accurate application to many-component natural silicate melts (ranging from primitive Fe/Mg-rich basalts to highly evolved silica-rich rhyolites), there are still opportunities for useful and accurate applications to natural systems with fewer components.
In particular, a similar modeling approach (using an alternate parameterization of the ideal associated solution model) has been successfully applied to few component systems like the CaO-MgO-Al$_2$O$_3$-SiO$_2$ (CMAS) system \citep{Ivanova2021,Shornikov2019}. 

The MELTS model on the other hand, which supports liquid calculations within VapoRock, permits explicitly non-ideal interactions between liquid components, representing them with quadratic excess energies.
The components of the model were selected to maximize the accuracy of the symmetric regular solution approximation, and the model was calibrated entirely on a wide range of natural magmatic compositions.
The quadratic interaction (Margules) parameters of the regular solution model were calibrated on 4666 statements of mineral-melt equilibria, 80\% of which were performed at 1~bar, the low-pressure conditions at which VapoRock is exclusively run.
As a demonstration of its accuracy for natural magma compositions, the MELTS model correctly predicts the saturation temperatures of silicate and oxide minerals typically to within $\pm$ 10~K relative to experimentally-determined temperatures \citep{Ghiorso1995}.  
Thus, these natural compositions lie comfortably within the compositional region of the MELTS calibration data, whereas the MAGMA liquid model relies entirely on compositional extrapolation from simplified binary and ternary subsystems \citep{Fegley1987, Schaefer2004}.  %while natural magmatic liquids represent compositional extrapolations for the MAGMA code, they fit nicely within the high-accuracy calibrated region of the MELTS model underlying VapoRock.

An additional fundamental challenge with modeling vaporizing silicate melts is the accurate prediction of liquid heat capacities.
This issue is especially important in vapor modeling given that experimental constraints on liquid properties span a wide range of temperatures (due to the strong composition-dependence of crystallization), and thus model applications often require significant thermal extrapolation to typical vaporization conditions.
It is well established experimentally that silicate liquid heat capacities generally remain nearly constant above the glass transition temperature (with the exception of certain compositions like pure silica) and generally follow compositional ideal mixing (again with the exception of quadratic cross-terms involving silica content) \citep{Richet1986, lange1992, DiGenova2014}.
Thus, experimental studies have found simple expressions that can accurately capture the behavior of silicate liquids to within a few percent and are built directly into the MELTS model \citep{Ghiorso1995}.
The IMCC model supporting MAGMA relies instead upon ideal mixtures of pseudo-species which individually possess fixed ideal-gas-like heat capacities.
As temperature increases, entropy favors ever-broader mixtures of species (each with their own enthalpies of formation that perturb the overall heat capacity of the mixture), ensuring that heat capacity is not held constant but changes with the evolving species populations.
Furthermore, the heat capacity value depends strongly on the particular set of pseudo-species in the model and is divorced entirely from direct measurements of silicate liquids.
As a consequence, while MELTS relies directly upon experimentally-determined expressions for liquid heat capacities, the IMCC model in MAGMA yields implied heat capacity values that cannot reproduce appropriate trends for either temperature or compositional dependence.

%Therefore, 
%While it is difficult to definitively demonstrate the absolute accuracy of the VapoRock code (due primarily to experimental limitations), VapoRock is clearly able to reproduce experimental data generally to within a factor of 3, and performs similar to (Ca, Al, \& Na), or better than (Si, Ti, K, Fe, \& Mg) the MAGMA code for terrestrial igneous rock compositions in complex systems. 
%While it is difficult to definitively demonstrate the absolute accuracy of the VapoRock code (due primarily to experimental limitations), 
VapoRock is clearly able to reproduce experimental data generally to within a factor of 3, and performs similar to (Ca, Al, \& Na) or better than (Ti, Fe, Mg, and especially K \& Si) the MAGMA code for terrestrial igneous rock compositions. 
In general, owing to the ability of MELTS to account for non-ideality of mixing by a regular solution model using pseudo-components, the activities it calculates are expected (from both theory and experiment) to be more precise than in the MAGMA code for complex natural silicate liquids. 
%Furthermore, these natural compositions lie comfortably within the compositional region of the MELTS calibration data, whereas the MAGMA liquid model relies entirely on extrapolation from simplified binary and ternary subsystems \citep{Fegley1987, Schaefer2004}.
%The natural magmatic compositions of primary concern to planetary science and geochemistry thus fall well inside the high-accuracy region of the VapoRock model, while the MAGMA model must utilize extrapolations from simplified subsystems that are both theoretically and experimentally unsupported. 
%\awnote{Add 1 paragraph summarizing the differences between MAGMA 04 and MAGMA 22. Emphasize how fundamental issues remain, and hence you cannot perfectly win. Focus on how K gets better but Si gets much worse. Also, what about heat capacity.}
We can further see that predictive accuracy is highly sensitive to changes in the MAGMA calibration.
The MAGMA 22 model was created to improve K-related predictions over the original version from 2004, but had the surprising side-effect of also dramatically worsening Si predictions (by roughly a factor of $\sim$10) for primitive to intermediate melts.
Most surprising is that the new calibration results in elemental values that diverge from those in VapoRock (by factors of 0.3 to 90 when normalised to Na; or 1 - 300 normalised to Ca) for highly evolved silica-rich liquids.
At its heart, these challenges all relate back to the liquid modeling framework and how accurately it captures compositional mixing for many-component systems.

\subsection{VapoRock limitations and ongoing developments}

The MELTS liquid model underlying VapoRock employs non-oxide components to better predict activities in compositional systems relevant to terrestrial magmatism (see Sec. \ref{sec:liquid_model}).
A drawback of this approach, however, is that the components do not allow for low \ch{SiO2} compositions to be accessed, as numerous non-oxide components are silica-based (e.g., \ch{KAlSiO4}, \ch{Fe2SiO4}, \ch{Na2SiO3} etc.). 
Hence, these elements (\ch{Fe^2+}, K, Na) cannot be represented without sufficient silica (and/or alumina) in the bulk composition. 
Therefore some exotic compositions, notably, a subset of Type B CAIs \citep[e.g.,][]{stolper1986} and indeed more extreme exoplanetary compositions \citep[e.g.,][]{Unterborn2020} with low \ch{SiO2} contents are inaccessible with VapoRock at present.
In future, the MELTS model could be extended to enable the calculation of compositions lying outside the current compositional system.

Although VapoRock returns partial pressures for Ti- and Cr-bearing gas species, caution is advised for the potential user.
MELTS only treats \ch{Ti^4+} and \ch{Cr^3+} in its solution model as the \ch{TiO2} and \ch{MgCr2O4} melt components, respectively.
This poses a potential problem for planetary mantles, which are thought to have largely formed in equilibrium with iron-nickel metal during core-formation \citep{frost2008, righter2006}, and thus experienced oxygen fugacities near or below the iron-wüstite buffer.
In particular, Cr is known to exist predominantly as \ch{Cr^2+} at high temperatures (rather than \ch{Cr^3+} as implicitly assumed by MELTS), with the \ch{Cr^2+}/\ch{Cr^3+} ratio approaching unity at $\Delta$IW=+3.5 and 1673~K in typical basaltic melts \citep{berry2006, berry2021}. 
Of all the multivalent cations, MELTS only directly represents multiple oxidation states for Fe, and subsumes any redox effects for Ti and Cr into the average behavior of liquids with Ti- or Cr-bearing compositions.
Therefore, partial pressures of Cr-bearing calculations performed at temperatures $>$~1673 K and oxygen fugacities below $\Delta$IW=+3.5 should be taken as indicative of the net average behavior of natural systems containing these elements.

A similar caveat applies to Ti-bearing gas species, though the range of temperatures and oxygen fugacities over which VapoRock is applicable is considerably wider.
This comes from the lower relative \fO2 at which the \ch{Ti^4+} to \ch{Ti^3+} transition takes place in basaltic melts.
\cite{leitzke2018} found no evidence for \ch{Ti^3+} in synthetic lunar mare basaltic glass equilibrated at $\Delta$IW=-2, though trivalent Ti does occur in some lunar pyroxenes \citep{simonsutton2018}.
VapoRock can therefore be used to faithfully predict Ti-bearing gas species down to $\Delta$IW=-2, but caution is advised at oxygen fugacities much below this, where predictions may begin to diverge from realistic behavior.

The compositional space in VapoRock comprises only metals and their associated oxides.
As such, H, C and other major volatile species are not included.
These elements readily bind with metal species as ligands, replacing or in conjunction with oxygen \citep[see][]{fegley2016, Sossi2018} to form additional species.
In steam atmospheres that can exert pressures of 10$^2$ to 10$^3$ bars on their planetary surface, associated gas species of major elements such as \ch{Si(OH)4} and \ch{Mg(OH)2} become more abundant than SiO and Mg, respectively \citep{fegley2016}. 
This occurs because the partial pressures of hydroxide species are related to the water fugacity of the atmosphere, by reactions of the type MO(\textit{s,l}) + \ch{H2O}(\textit{g}) = \ch{M(OH)2}(\textit{g}) (where M = a divalent metal cation).
Consequently, the hydroxide species become predominant as \textit{f}\ch{H2O} increases.
As noted by \citet{fegley2016} and \citet{Sossi2018}, some elements (in particular the alkalis) are sensitive to the halogen content of the atmosphere through reactions of the form A(\textit{g}) + 0.5\ch{Cl2}(\textit{g}) = ACl(\textit{g}), where A represents an alkali metal \citep[see][for an extension of the MAGMA model that includes Cl complexation]{Tian2019}. 
Because there are 1.5 moles of gas in the reactants and 1 mole in the products, \textit{p}(ACl)/\textit{p}(A) ratio is proportional to (\textit{P}$_T$)$^{0.5}$, where \textit{P}$_T$ is the total pressure of the atmosphere.
In planetary scenarios, the \textit{P}$_T$ may exceed 1 bar, promoting the formation of these species.
Clearly, a self-consistent understanding of the speciation of atmospheres surrounding rocky planets is a complex question, for which VapoRock, at present, is best suited to modelling nominally anhydrous, low-pressure examples.
Future work must be done to incorporate volatiles into the VapoRock modeling framework and to assess the accuracy of such a volatile-bearing model.

%\subsection{Basic Planetary Vaporization Simulation}
%

%\begin{figure}[!ht]
%    \centering
%    \includegraphics[width=1\textwidth]{figs/sns_multiplot_Delano1986_JANAF.pdf}
%    \caption{\njnote{Example: }VapoRock partial pressures for the 12 most common gas species for melts with green and orange Lunar glass compositions \citep{delano1986}.}
%    \label{fig:logP-Delano1986}
%\end{figure}
%
%\begin{figure}[!ht]
%    \centering
%    \includegraphics[width=1\textwidth]{figs/sns_multiplot_grn_species_JANAF.pdf}
%    \caption{\njnote{Alternative, also available for orange ('org'): }VapoRock partial pressures for all gas species for the green Lunar glass composition \citep{delano1986}.}
%    \label{fig:logP-grn-Delano1986}
%\end{figure}

\subsection{Vaporization of the Bulk Silicate Earth}

\begin{figure}[!ht]
    \centering
    \includegraphics[width=1\textwidth]{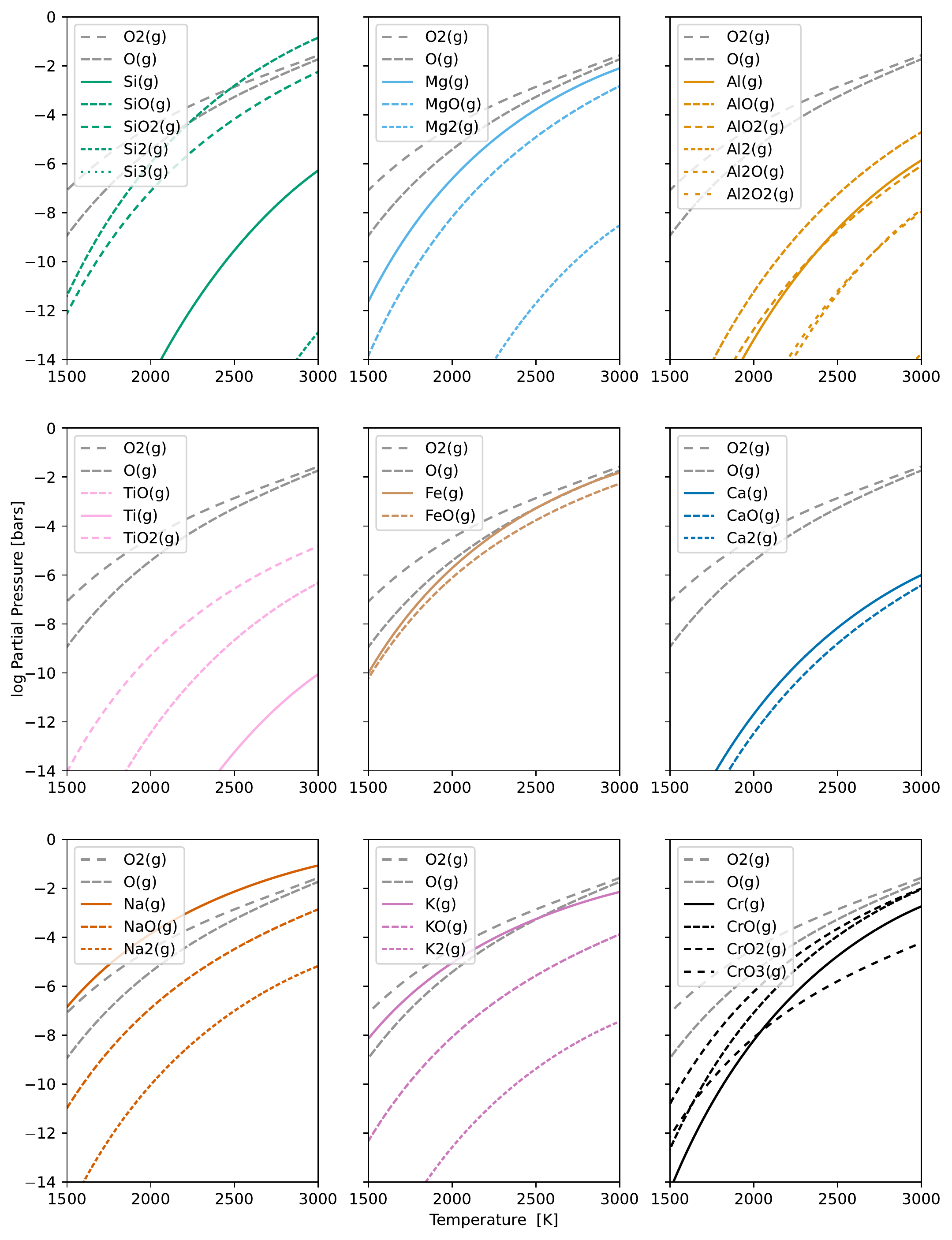}
    \caption{Predicted gas species partial pressures for vaporization of Bulk Silicate Earth (BSE) liquid calculated using VapoRock. Thermal trends are separated according to cation type and use an expanded color-coding and line-style convention consistent with Fig.~\ref{fig:Pfrac-Schaefer}. BSE composition and oxygen fugacity trend with temperature are taken from \cite{Schaefer2009} and \cite{oneill1998}, respectively.}
    \label{fig:logP-BSE-Schaefer2009}
\end{figure}

The Bulk Silicate Earth (BSE) represents an estimate for the composition of the Earth minus its iron-nickel core, calculated by combining geological and cosmochemical constraints \citep{palmeoneill2014}.
To explore the outgassing of molten terrestrial mantle materials after significant volatile loss, we model the vaporization of the molten BSE without major volatiles (H, C, N, S).
The partial pressures of the equilibrium atmosphere above the BSE are calculated over a temperature range from 1500 to 3000 K, at an \fO2 given by \cite{Schaefer2009} for congruent vaporization, and ranges from approximately $\Delta$IW=+4.5 at 1500 K to $\Delta$IW=+1 at 3000 K (Fig. \ref{fig:logP-BSE-Schaefer2009}).
%\psnote{@Noah, can you provide a table in the supplementary information with the output for the BSE - i.e., including fO2 and the partial pressures as a function of temperature?}
As per the other compositions listed in Table \ref{tab:MAGMA_comps}, Na(\textit{g}) is the predominant species in the vapour until $\sim$2700 K, at which point it is overtaken by SiO(\textit{g}).
This basic behavior was also seen in previously published calculations using VapoRock for a variety of other natural liquid compositions explored in \citet{Jaeggi2021}.
The major gas species for each element is Mg(\textit{g}), AlO(\textit{g}), \ch{TiO2}(\textit{g}), Fe(\textit{g}), Ca(\textit{g}), K(\textit{g}) and \ch{CrO2}(\textit{g}).\\

Most elements therefore vaporize via reactions in which the gaseous species is more reduced than in the liquid \citep[cf.]{sossi2019evaporation}.
In terms of number of electrons transferred (\textit{n}), the dominant reactions for Si, Mg, Al, Fe and Ca involve \textit{n} = 2, such as Fe$^{2+}$O(\textit{l}) = \ch{Fe^0}(\textit{g}) + 0.5\ch{O2}(\textit{g}).
Even though the dominant species are also monatomic gases, the alkalis evaporate according to \textit{n} = 1 reactions, namely, A$^{+}$O$_{0.5}$(\textit{g}) = A$^{0}$(\textit{g}) + 0.25\ch{O2}(\textit{g}). 
Titanium differs in that its speciation in the melt is identical to that in the vapour, \ch{TiO2}, such that its partial pressure is independent of \fO2. 
At more reducing conditions, TiO(\textit{g}) becomes more significant relative to \ch{TiO2}, proportional to \fO2$^{-0.5}$.
Of the elements studied here, chromium is unique in that it is the only element whose most stable gas species, \ch{Cr^4+ O2} is more oxidised than in the silicate liquid \cite[\ch{Cr2^3+ O3}, see also][]{sossi2019evaporation}.
More oxidizing conditions therefore promote the vaporization of Cr, as opposed to all other major elements whose volatilities decrease with increasing \fO2.

\subsection{Constraining oxygen fugacity during vaporization}
As highlighted for the BSE, the oxygen fugacity during vaporization of silicate melts exerts strong controls on vapor abundances, since \ch{O2} appears as either a reactant or product in most liquid vaporization reactions (assuming oxides are used as system components).
And yet, due to its low partial pressure in most situations, determining oxygen fugacity in either experiments or natural systems poses problems.
Though some experiments do measure and report molecular (or atomic) oxygen abundances, these %are only measurable over specific temperature intervals 
often do not cover the full range of the experiments and shift depending on the composition of the sample (which itself can change due to fractional vaporization). 
Indeed, owing to the high background levels of \ch{O2} relative to other molecular species, most KEMS studies no longer determine \fO2 directly, but rather calculate it assuming congruent evaporation of other components \citep[see][their Eq. 8]{kobertz2014}.

% In accordance with the dominant role played by the liquid in the outgassing regime, the vaporization process is fully determined by the state of the liquid.
%Within the outgassing regime, the state of the liquid fully determines the vaporization process.
%The liquid's composition (given in terms of weight-percent oxides) sets the chemical potentials of the system.
%Extra attention must be given to the oxygen abundance (or fO2) of the system, since imperceptibly small shifts in oxygen (at less than the parts-per-billion level) \awnote{ppb, ppm, .1\%, which one??} can dramatically affect the abundances of outgassed species.
%Silicate liquids are capable of adopting the full range of oxygen fugacities observed in nature, from extremely reducing (in equilibrium with metallic iron) to fully oxidized (in equilibrium with pure oxygen).
%For non-highly reduced states, this is generally accommodated through the balance of FeO and Fe2O3, but even extremely reduced systems with negligible ferric iron can equilibrate with any desired O2 abundance through minute shifts in the amount of dissolved oxygen.
Most existing codes (including MAGMA) adopt a deterministic approach to calculating \fO2, using cation/oxygen abundance ratio constraints to predict the outgassed \ch{O2} partial pressure from a magma \citep[e.g.][]{Fegley1987}.
This is achieved by iteration, adjusting the \fO2 until it yields a total cation/oxygen ratio in the vapor that matches the liquid.
%is equal to the sum of the number of moles of oxygen associated with the congruent vaporization reaction(s) of each melt oxide component.
These models assume, therefore, that the oxygen budget of the condensed phase dominates that in the gas phase.
%However, due to the low oxygen abundances in the vapor, its budget is liable to change by processes such as removal by escape or deposition, molecular processing, or inflow of highly reducing nebular \ch{H2} gas. %should be able to dramatically alter the availability of oxygen.
%Becasuse vaporisation stoichiometries are well-defined, 
While this approach has produced favorable comparisons with experiments \citep{Schaefer2004}, it relies on the ad-hoc constraint of balancing cation/oxygen abundances, which does not accurately reflect thermodynamic equilibrium for non-congruent vaporization.%
\footnote{In congruent vaporization, the liquid and gas compositions are identical, as atoms or molecules in the liquid are vaporized indiscriminately. This type of phase change occurs in nature but generally does not hold true for silicate liquids except for simple endmember compositions.}
In reality, the ferric/ferrous iron ratio of the liquid, (as the dominant redox-sensitive element), sets the \fO2 of the volumetrically minor gas phase, thus allowing distinct cation/oxygen ratios for the liquid and coexisting vapor.
%This ratio is often not known a priori, however.
Moreover, \fO2 in natural settings may be buffered by another phase (e.g., Fe-Ni metal) that is not explicitly considered in the model system (liquid+vapor). 
%Even in cases where the redox state of the condensed phase does dictate the oxygen fugacity,
The \fO2 of nominally `fully-reduced' compositions (with no measurable ferric iron) are not fully defined by the liquid composition alone.
Thus additional constraints on oxygen abundance must be imposed to estimate and predict the evolving oxygen fugacity during vaporization of highly reduced materials like primitive planetary mantles.%
\footnote{Such situations highlight the advantages of retaining \fO2 as an input of the model, rather than a single predicted outcome, since many realistic systems remain open with respect to oxygen or experience additional processes that exert dominant control over the \ch{O2} budget.}
This issue of predicting self-consistent oxygen vaporization abundances is a rich and complex topic that will be addressed in a future study.
%But it is clear that the oxygen abundance ratio method runs into serious problems, given that it always deterministically yields a single 'preferred' fO2, in spite of the reality that the fO2 of silicate liquids is determined by their equilibrium conditions, and can take on any value seen in nature.
%This issue is a particularly nuanced one that is beyond the scope of the current investigation and will be explored in detail in future work.
To retain complete generality, \fO2 is thus an input parameter of the VapoRock model.
For convenience, this \fO2 level can be set using either its numerical unitless value, $\log_{10}$(\fO2 / 1 bar), or it can be described relative to a planetary redox buffer (e.g. IW, QFM, NNO).
%Some caution should be applied, however, to interpreting this buffer-relative value.
%There is some evidence that the thermal trends of common redox-buffers may begin to deviate from their commonly parameterized paths at very high temperatures relevant to vaporization.
%In such cases, we recommend reporting both the buffered values as well as the raw log(fO2) values.
%Further investigation into this issue will be deferred to future publications.

%Beyond constraining experimental conditions based on empirical measurements, building a true understanding of the underlying factors that govern evolving oxygen abundances in natural systems poses even greater challenges.
%Much of the published modeling work \citep[see work of Fegley and Schaefer, e.g.][]{Fegley1987,Schaefer2004} proceeds as if the samples themselves possess unique intrinsic oxygen fugacities that are well constrained at different vaporization temperatures.

%\awnote{Actually Kress and Carmichael is probably stupidly accurate for these conditions, may suffer somewhat at very reducing conditions, but it is by far the best we've got and way more accurate than the congruent vaporization approach.}
%it is unclear whether existing models that relate \ch{Fe^3+}/\ch{Fe^2+} ratios in silicate melts to \fO2  \citep[e.g.][]{Kress1991} are sufficiently accurate to predict the evolving oxygen fugacity during vaporization of these materials. 

\subsection{Constraining oxygen fugacity in planetary atmospheres}

\begin{figure}[!ht]
    \centering
    \includegraphics[width=\textwidth]{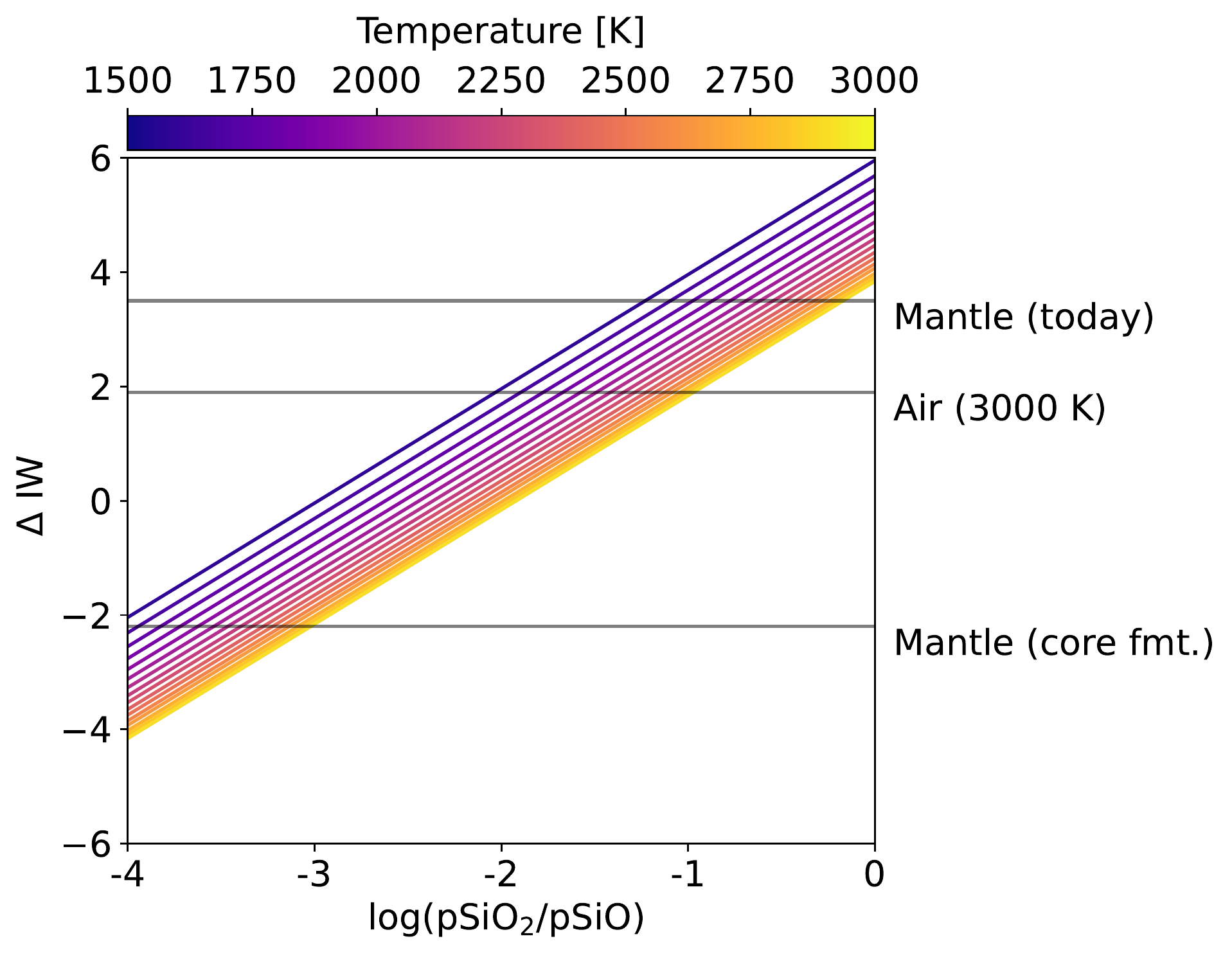}
    \caption{Oxygen fugacity trends as a function of the log(\textit{p}\ch{SiO2}/\textit{p}SiO) ratio of the atmosphere, expressed as log unit deviations relative to the iron-wüstite buffer. Lines represent different isotherms ranging from 1500~K to 3000~K. Horizontal lines show reference oxygen fugacity for air (at 3000~K) and the Earth's mantle (today and during core formation). Earth mantle oxygen fugacity data from \cite{frost2008}. }
    \label{fig:SiO-SiO2}
\end{figure}

From an atmospheric standpoint, the oxygen fugacity can be inferred from the relative abundances of coupled vapor species with variable oxidation states (like \ch{TiO2} and \ch{TiO}) with the help of a thermochemical database of vapor species, though this calculation is rarely presented in most experimental studies.
Calculations of the evaporation of silicate materials indicate, unsurprisingly, that SiO(\textit{g}) is one of the prevailing species throughout the temperature range 1500 - 3000 K, reaching 0.01 to 0.1 bar at the upper end of this range.
It has a sister species, \ch{SiO2}(\textit{g}), that is typically present at levels 10 - 100 times lower than SiO(\textit{g}) (cf. Fig. \ref{fig:logP-BSE-Schaefer2009}). 
At equilibrium, the partial pressures of the two species are related by the oxygen fugacity:
\begin{equation}
    \ch{SiO}(g) + 1/2\ch{O2}(g)  = \ch{SiO2}(g)
    \label{eq:SiO-SiO2}
\end{equation}
Applying the law of mass action for an ideal gas and solving for \fO2 yields:
\begin{equation}
    f\ch{O2} = (p\ch{SiO2}/[p\ch{SiO} \times K_{(\ref{eq:SiO-SiO2})}])^2
    \label{eq:pO2-Si}
\end{equation}
An expression for the equilibrium constant of equation \ref{eq:SiO-SiO2}, K$_{(\ref{eq:SiO-SiO2})}$, is given by tabulated data from JANAF, as implemented in VapoRock:
\begin{equation}
    logK_{(\ref{eq:SiO-SiO2})} = 10278/T - 4.0436
    \label{eq:logKSiSiO}
\end{equation}
These series of equations can be used to predict the \fO2 of a given atmosphere provided the \textit{p}SiO, \textit{p}\ch{SiO2} and temperature are known, and chemical equilibrium is assumed.
A grid of \fO2 values as a function of these variables is shown in Fig. \ref{fig:SiO-SiO2}.\\

\cite{zilinskas2022} modelled predicted spectral responses from the emission of fictive silicate atmospheres around rocky exoplanets, including 55-Cnc-e, among others. 
These authors demonstrated that SiO and \ch{SiO2} have strong characteristic features, at 9 $\mu$m and 7 $\mu$m, respectively, which are readily observable with the MIRI instrument of JWST.
Moreover, the broad feature produced by SiO is expressed as an emission band, whereas \ch{SiO2} is a strong absorber, giving rise to an acute contrast in the mid-infrared as a result of the presence of the two molecules.
Our comparative benchmark of the VapoRock and MAGMA codes for natural silicate melts (see Table \ref{tab:MAGMA_comparison}) shows that MAGMA systematically underestimates Si vapor abundances by a factor of 2 to 300 (depending on liquid composition and the model version of MAGMA), and thus these findings represent lower bounds indicating that Si-species spectrometry is likely even more readily observable than reported in \citet{zilinskas2022}.
In cases of extreme vaporization, like the anhydrous compositions explored by \cite{zilinskas2022}, Si-containing vapor species provide a unique opportunity for directly measuring the oxygen abundances within planetary mantles.\\

Fig.~\ref{fig:SiO-SiO2} illustrates that the temperature dependence of reaction \ref{eq:SiO-SiO2} is weak relative to the expected range of \fO2 variation. Consequently the \textit{p}\ch{SiO2}/\textit{p}SiO, if determined accurately together with temperature, is a sensitive probe of oxygen fugacity of the atmosphere.
Given the slope of log\fO2 = 2log(\textit{p}\ch{SiO2}/\textit{p}SiO), an uncertainty of $\pm$~1 log unit in (\textit{p}\ch{SiO2}/\textit{p}SiO) corresponds to an uncertainty of $\pm$~2 log units in \fO2 at constant temperature.
This level of precision would be sufficient to distinguish (in terms of oxygen fugacity) between an atmosphere similar to that of Earth during core formation ($\Delta$IW=-2.2) from its present-day mantle value ($\Delta$IW=+3.5; see Fig. \ref{fig:SiO-SiO2}; \cite{frost2008}) and therefore would be able to track large-scale transitions in the oxidation state of planetary atmospheres.

This conclusion does come with several caveats.
As noted, VapoRock does not currently account for H-bearing species, that may overwhelm any IR fingerprint coming from SiO and \ch{SiO2}. 
However, the presence of volatiles like H or C provide ample opportunity for spectral retrieval of atmospheric oxygen abundances.
The assumption of chemical equilibrium may also break down due to photochemical reactions induced by UV absorption at wavelengths shorter than 300 nm \citep[see][and references therein]{Schaefer2009} in the upper regions of the atmosphere.
For the measurement of \textit{p}\ch{SiO2}/\textit{p}SiO to be related to that of the magma ocean, a full atmospheric radiative transfer model is required to map the distribution of the two species as a function of the atmosphere's scale height \citep{Ito2015, zilinskas2022}. 
Nevertheless, we propose that the SiO/\ch{SiO2} ratios of rocky planetary atmospheres are most promising proxies of \fO2, and, consequently, the interior redox state of their mantles.

%\subsection{Comparing VapoRock with Slag vaporization experiments}
% <!-- ## [[202101200609]] Comparing VapoRock with Slag vaporization experiments -->

\section{Conclusion}
We develop and fully validate a new open-source code for silicate liquid vaporization, VapoRock, relevant for modeling outgassing of silicate magma oceans and volcanic venting in the volatile-free Si-Mg-Fe-Al-Ca-Na-K-Ti-Cr-O system.
This code combines descriptions of the thermodynamic properties of vapor species from the NIST-JANAF \citep{Chase1998} thermochemical tables---and alternative expressions of Lamoreaux and coworkers \citep[][] {Lamoreaux1984,Lamoreaux1987}---with the silicate liquid model of MELTS \citep{Ghiorso1995}.
%The model accurately predicts vapor abundances for a wide range of natural magmatic compositions .
Vapor pressures and associated melt oxide species activities calculated by VapoRock are shown to reproduce experimental results for vaporization of complex synthetic and natural silicate liquids to within a factor of $\sim$3 (consistent with experimental uncertainties) for quantities that vary over 6~to~10 orders of magnitude.
Furthermore, we carry out an extensive comparison of calculated vapor pressures using VapoRock and two versions of the community-standard benchmark code MAGMA \citep{Fegley1987,Schaefer2004,Jiang_etal2019}, for a wide range of natural magma compositions ranging from dunites to rhyolites.
We find that VapoRock yields vapor pressures that are on par with (Ca, Al, \& Na) or more accurate than (Mg, Ti, Fe, and especially K, Si) the MAGMA code.
In particular, the newest version of MAGMA (2022) suffers from underestimating Si outgassing for all compositions studied, and this effect is amplified for silica-rich liquids (like the rhyolitic Bishop Tuff) where errors increase substantially for nearly every cation in the gas.
These systematic differences (which are as large as 2-3 orders of magnitude for K and Si, depending on the version of MAGMA used) are generally attributable to fundamental distinctions in the underlying liquid models and their ability to capture non-ideal interactions for realistic natural silicate melts.
The model is used to explore anhydrous vaporization of a bulk silicate Earth composition representative of rocky planetary mantles, showing that the major species are Na(\textit{g}) below 2700 K and SiO(\textit{g}) above it \citep[confirming observations of a previous application of VapoRock in][]{Jaeggi2021}. 
We additionally explore the possibility of using the ratio of silica vapor species, SiO and SiO$_2$, as a spectroscopic proxy for inferring the oxidation state of mantle-outgassed exoplanetary atmospheres.
It is shown that a precision of $\pm$~1 log unit on SiO/SiO$_2$ would be sufficient to discern between different evolutionary stages experienced over Earth's history and, by extension, those of exoplanets.

\begin{acknowledgements}

The authors are indebted to multiple agencies for their generous financial support of this work.
ASW thanks the National Science Foundation (EAR 1725025), NASA (80NM0018D0004), and the Turner Postdoctoral Fellowship for grant funding that made this research possible.
Financial support has been provided to NJ by the Swiss National Science Foundation Fund (200021L\_182771/1). PAS thanks the Swiss National Science Foundation (SNSF) via an Ambizione Fellowship (180025) and an Eccellenza Professorship (203668) and the Swiss State Secretariat for Education, Research and Innovation (SERI) under contract number MB22.00033, a SERI-funded ERC Starting Grant `2ATMO'.  DJB acknowledges SNSF Ambizione Grant 173992 and support from a CSH Fellowship at the University of Bern.
We additionally thank Laura Schaefer and an anonymous reviewer for constructive comments that improved the manuscript, Mark Ghiorso for many useful conversations over the course of this work, Bruce Fegley for sharing the latest version of the MAGMA code (October, 2022), as well as Fabian Seidler for the python wrappers he provided for MAGMA.
\end{acknowledgements}

\appendix

\section{Shomate Equation}
\label{app:shomate-eqns}
To evaluate the chemical potential of each vapor species, the ideal gas approximation must be coupled with an analytic expression for energy as a function of temperature, like the Shomate equation used by the thermochemical tables of JANAF, \cite{Lamoreaux1984} and \cite{Lamoreaux1987}.
% <!-- ## Shomate Equation -->
% <!-- # [[202006090659]] Shomate Equation -->
% <!-- The Shomate equation provides an analytic empirical description of the chemical potential of a phase as a function of temperature evaluated at a (1 bar) reference pressure. -->
The Shomate equation empirically captures energy variations for a wide selection of materials over a broad temperature range, and multiple piecewise models are sometimes combined to retain desired accuracy over 1000+ degree intervals.
The generalized polynomial form of the Shomate equation describes molar enthalpy in kJ/mol:
\begin{equation}
    \begin{split}
    \Delta \bar{H}^0 &= \bar{H}^0 - \bar{H}^0_{298.15} \\
    &= At + \frac{B}{2} t^2 + \frac{C}{3} t^3
    + \frac{D}{4} t^4 - E t^{-1} + F
    \label{app:eq:shomate_enthalpy}
    \end{split}
\end{equation}
molar entropy in J/mol/K:
\begin{equation}
    \bar{S}^0 = A\ln{t} + Bt + \frac{C}{2} t^2 + \frac{D}{3} t^3 - \frac{E}{2} t^{-2} + G
    \label{app:eq:shomate_entropy}
\end{equation}
and molar heat capacity in J/mol/K:
\begin{equation}
    C^0_P = A + Bt + C t^2 + D t^3 + E t^{-2}
    \label{app:eq:shomate_heat_cap}
\end{equation}
where $t$ is temperature given in kilo-Kelvin ($t = T/1000$).
These are combined to evaluate the molar Gibbs energy (or chemical potential):
\begin{equation}
    \bar{G}^0 \equiv \mu^0 = \Delta \bar{H}^0 - T \bar{S}^0
    \label{app:eq:shomate_Gibbs}
\end{equation}
These numerical coefficients ($A,B,C,D,E,F,G$) are provided for a huge set of phases in the JANAF thermochemical database, as well as the tables of \cite{Lamoreaux1984} and \cite{Lamoreaux1987}.
Using these expressions, we can rapidly evaluate the chemical potential reference state for every vapor species in the system, allowing us to determine the equilibrium partial pressure of each one.

\section{Stoichiometric conversion between liquid and system components}
\label{app:stoic}

To carry out equilibrium liquid/vapor computations with our thermochemical database, we first convert both liquid components and vapor species into their equivalent representation using system components (oxides) with additional oxygen (vapor) as needed to ensure stoichiometric balance.
%  <!-- [[202101111354]] -->
Some representative examples are given for a number of liquid components and vapor species of varying complexity:
\begin{equation}
\begin{aligned}
\textbf{liquid components}\\
\ch{Fe2O3}(\ell) &= \ch{Fe2O3}(c)\\
\ch{Fe2SiO4}(\ell) &= 2\,\ch{FeO}(c) + \ch{SiO2}(c)\\
\ch{KAlSiO4}(\ell) &= \frac12 \ch{K2O}(c) + \frac12 \ch{Al2O3}(c) + \ch{SiO2}(c)\\
\textbf{vapor species}\\
\ch{Na2O}(v) &= \ch{Na2O}(c)\\
\ch{NaO}(v) &= \frac12 \ch{Na2O}(c) + \frac14 \ch{O2(}v)\\
\ch{Si2}(v) &= 2\,\ch{SiO2}(c) - 2\,\ch{O2}(v)\\
\ch{Al2O2}(v) &= \ch{Al2O3}(c) - \frac12 \ch{O2}(v)\\
\end{aligned}
\label{app:eq:conversion_example}
\end{equation}
where chemical compounds exist in the liquid ($\ell$) or vapor ($v$) phase are represented using system components ($c$), typically oxides.
Due to the linearity of chemical potentials, this allows us to work in a common basis of oxides while interfacing between the condensed and vaporized portions of the system.

Generalizing the specific examples given, compositional conversion of the liquid endmembers is determined by stoichiometry:
%  <!-- [[202101121125]] -->
\begin{equation}
    \phi_i^\ell = \sum_{j} \nu_{ij}^\ell c_j
    \label{eq:comp_conversion}
\end{equation}
where $\phi_i^\ell$ is the $i^{th}$ endmember component for the liquid phase, $c_j$ is the vector of basic system components (like oxides), and $\nu_{ij}^\ell$ is the stoichiometry matrix for the liquid phase reflecting the composition of the $i^{th}$ endmember in terms of the $j^{th}$ system component.
Given the linearity of chemical potentials, the relation between phase and system chemical potentials can be written:
\begin{equation}
    \mu_{i}^{\ell} = \sum_j \nu_{ij}^{\ell} \mu_j
    \label{eq:phaseVSchem_pot}
\end{equation}
which is inverted to obtain the expression needed for thermodynamic computation:
\begin{equation}
    \mu_j = \sum_i \tilde{\nu}^\ell_{ji} \mu_i^\ell
    \label{eq:phaseVSchem_pot_inverted}
\end{equation}
where $\tilde{\nu}^\ell = ({\nu^\ell})^{-1}$ is the inverse stoichiometry matrix for the liquid phase.
Thus we obtain the chemical potentials of the system in terms of the desired (oxide) components through a simple dot product with the inverse liquid stoichiometry matrix.
Having brought both the liquid endmember and vapor species into a common basis of system components, we are able to meaningfully combine the thermochemical data on each phase to preform useful computations.

\section{Calculating partial pressures from the law of mass action}
\label{app:mass-action-law}

In order to determine the hypothetical partial pressure of the $i^{th}$ gas species in an idealized system, given a composition for the liquid and an oxygen fugacity (\fO2), the law of mass action is used.
A common example is the congruent dissociative evaporation of SiO$_2$(\textit{l}) into its gaseous constituents, SiO(\textit{g}) and O$_2$(\textit{g}).
The reaction takes the form:
\begin{equation}
    SiO_2(l) = SiO(g) + 0.5\,O_2(g)
    \label{eq:SiO2_evap}
\end{equation}

The equilibrium constant, \textit{K} for this reaction (assuming \textit{p} = \textit{f}) is:
\begin{equation}
     K= p(SiO)\,p(O_2)^{0.5}/a(SiO_2)
    \label{eq:SiO2_evap_K}
\end{equation}

By the law of mass action, the equilibrium constant can be related to the Gibbs Free Energy change of the reaction by:

\begin{equation}
     \Delta G_{rxn} = -R\,T\,\ln K
    \label{eq:SiO2_evap_delG}
\end{equation}

Where $\Delta G_{rxn}$ is given by:

\begin{equation}
     \Delta G_{rxn} = G(SiO) + 0.5\,G(O_2) - G(SiO_2)
    \label{eq:SiO2_evap_delG-G}
\end{equation}

Values of the Gibbs energies of the compounds, G, can be compiled from lookup tables such as in JANAF (as Gibbs Energies of formation from the elements, $\Delta _f$G$^{\circ}$), or calculated using the Shomate equations. 
Equation \ref{eq:SiO2_evap_K} is re-arranged to solve for \textit{p}(SiO) and evaluated over the desired temperature range. 
In order to recover \textit{p}(SiO), \textit{p}(O$_2$) must also be known. 
For the idealized system, \textit{a}(SiO$_2$) = \textit{x}(SiO$_2$), yielding \textit{p}(SiO)$^{0}$ that can be compared with experimental data or VapoRock outputs.

\section{Thermodynamics of Silicate Liquid Mixtures}
\label{app:mixed-silicate-thermo}
The thermodynamic properties of silicate melts are manifestations of the atomic structure of the liquid.
The physical state of the liquid is described by the overall degree of polymerization, depicting how interconnected the strong silica network is within the liquid's atomic pseudo-lattice, as well as the population of oxygen-bonded species for each metal cation \citep[e.g.][]{Mysen2018}.
These properties describe how `active' a given element or species is within the silicate liquid, and therefore influences its activity coefficient.

In studies of silica-metal oxide liquid binaries \citep[e.g.][]{charles1967activities, ryerson1985oxide}, the nature and mole fraction of the metal oxide influences the activity of silica. 
The activity coefficient, $\gamma$SiO$_2$, exceeds 1 for alkaline earth oxides (e.g. Mg and Ca) and decreases below 1 for alkali oxides (e.g. Na and K).
The magnitude of the effect is proportional to the cation's ability to disrupt (or strengthen) the silicate melt network, which can be estimated by its Z/r (charge/ionic radius) ratio.
For example, the silica activity coefficient ($\gamma$SiO$_2$) experiences larger negative deviations from ideality due to \ch{K2O} than \ch{Li2O}, while the largest positive deviations from ideality for alkaline earth cations decrease in the order Mg $>$ Ca $>$ Sr $>$ Ba, reflecting differences in cation size.

These effects are not composition-independent, however, since each metal cation has a different energetic preference for particular sites in the liquid pseudo-lattice, enabling higher order mixtures to deviate significantly from behavior along simple binaries \citep{ryerson1985oxide}.
%causing mixed ternary and higher order liquids to possess different enthalpies of formation than those implied by the individual binaries.
%This effects is often referred to as 
The `mixed alkali effect'---which applies to all network-modifying and network-forming cations to varying degrees (e.g. Na, K, Mg, Ca, Fe, Ti, P)---describes how intermediate liquid mixtures involving multiple metal oxides are energetically stabilized or disrupted far beyond expectations from simple silica metal-oxide binaries \citep[e.g.][]{Mysen1990}.
A clear illustration of this effect can be seen in the simplified \ch{NaAlSiO4}--\ch{KAlSiO4}--\ch{SiO2} liquid system, where the heat of mixing describing Na-K exchange becomes increasingly negative and asymmetrically biased toward Na as the silica content is reduced, strongly favoring mixed Na-enriched compositions with a maximum depth to the energy well appearing at Al/(Al+Si)$\approx$1/3 and Na/(Na+K)$\approx$1/4 \citep[see discussion in][]{Mysen1990}.
In the absence of pseudo-species that adequately sample these intermediate compositions, the pronounced asymmetry and energetic magnitude of this mixing effect violates the assumptions of ideal mixing embedded in the IMCC model underpinning the MAGMA code.
The end result is that silicate liquid models built upon simplified subsystems (which cannot possibly account for the mixed alkali effect) improperly describe the stabilities of (or energetic preference for) intermediate mixed compositions.

\bibliography{references.bib}{}
\bibliographystyle{aasjournal}

\end{document}

\title{Supplement: VapoRock: Thermodynamics of vaporized silicate melts for modeling volcanic outgassing and magma ocean atmospheres}

\correspondingauthor{Aaron Wolf}
\email{aswolf@umich.edu}
\author[0000-0003-2415-0508]{Aaron S. Wolf}
\affil{Earth and Environmental Sciences, University of Michigan, 1100 North University Avenue, Ann Arbor, MI 48109-1005, USA}
\author[0000-0002-2740-7965]{Noah J\"aggi}
\affil{Physics Institute, University of Bern, Sidlerstrasse 5, 3012 Bern, Switzerland}
\author[0000-0002-1462-1882]{Paolo A. Sossi}
\affil{Institute of Geochemistry and Petrology, Department of Earth Sciences, ETH Zurich, Clausiusstrasse 25, 8092 Zurich, Switzerland}
\author[0000-0002-0673-4860]{Dan J. Bower}
\affil{Center for Space and Habitability, University of Bern, Gesellschaftsstrasse 6, 3012 Bern, Switzerland}
%\author[0000-0003-2425-3793]{Andr\'e Galli}
%\affiliation{Physics Institute, University of Bern, Sidlerstrasse 5, 3012 Bern, Switzerland}

\begin{abstract}
This document contains supplementary figures and discussion for "VapoRock: Thermodynamics of vaporized silicate  \& melts for modeling volcanic outgassing and magma ocean atmospheres".

%Data Guide of AAS Journals: "As with source codes, the AAS Journals are strongly encouraging authors to place related content in a persistent repository and to link to that material through a DOI link. In the case that authors choose to not use such archives then all of the files can be packaged together and submitted as a UNIX tar file. Again, a metadata header should then be included in the packaged file as a separate file called ReadMe."

% ReadMe should read: 
% Title:
% Authors:
% 
% Description of contents: ## Please replace this with description of the tar package that explains for the % reader what types of files are included and how to use them.
% 
% System requirements: ## Please indicate if the files require any special programs (e.g. IRAF) to be used. ##
% 
% Additional comments: ## Please replace this with additional comments or even text from the main paper useful % for the reader. ##

\end{abstract}

\section{VapoRock partial pressures of BSE composition based on Lamoreaux data}

\begin{figure}[ht]
    \centering
    \includegraphics[width=1\textwidth]{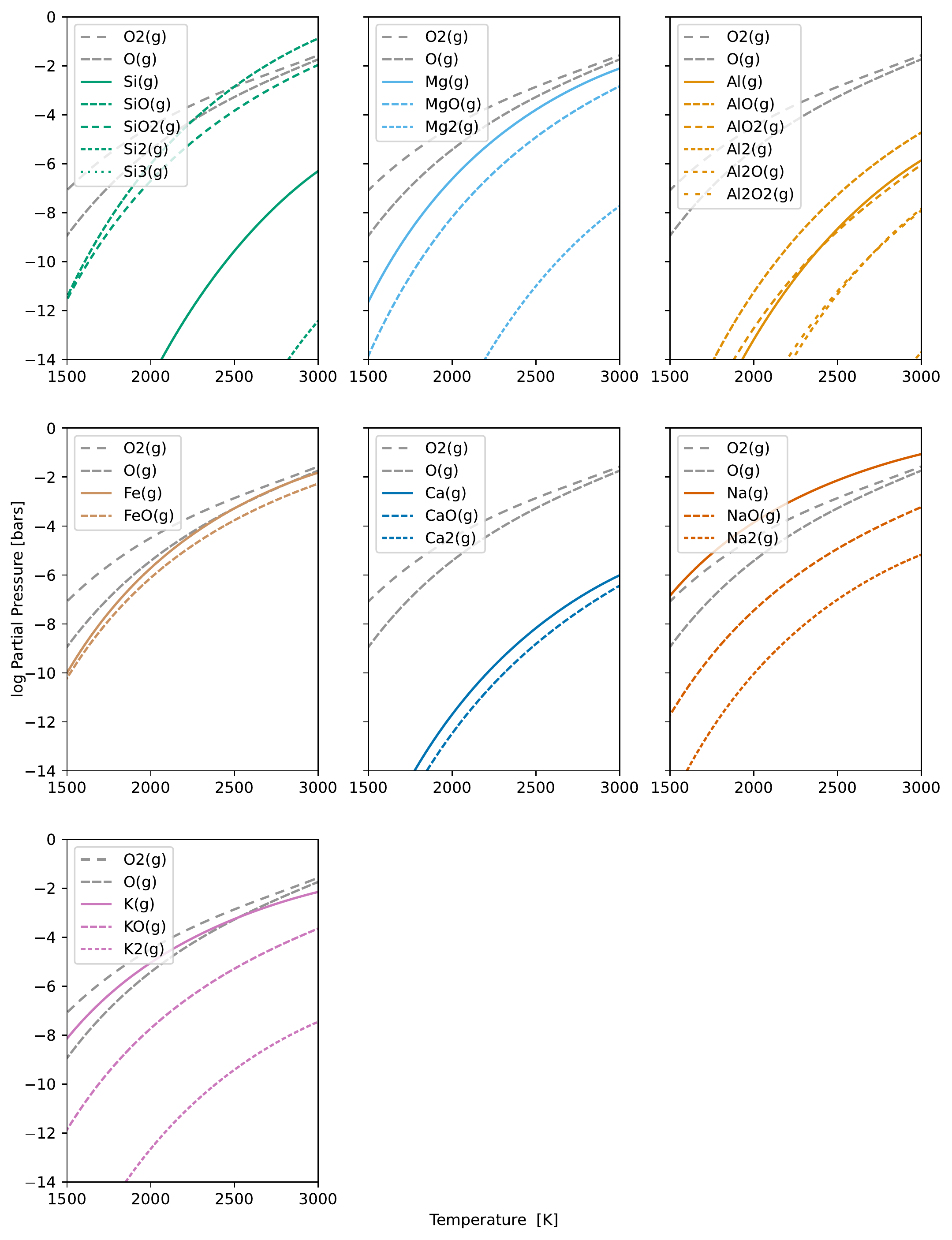}
    \caption{VapoRock partial pressures for all components separated by species. BSE composition and oxygen fugacity as a function of temperature was taken from \cite{Schaefer2009} and \cite{oneill1998} respectively. The thermochemical data of Lamoreaux and colleagues \citep[][]{Lamoreaux1984,Lamoreaux1987} was used. Cr and Ti species are absent, as they are not included in the original database. Fe thermochemical data was added from NIST-JANAF data \citep{Chase1998}. \label{fig:logP-BSE-Schaefer2009-Lamoreaux}}
\end{figure}

VapoRock vapor pressures for BSE composition based on Lamoreaux thermochemical database \citep[][]{Lamoreaux1984,Lamoreaux1987}. Figure~\ref{fig:logP-BSE-Schaefer2009-Lamoreaux} thereby recreates Fig.~4 in the parent manuscript. Cr and Ti species are absent, as they are not included in the original database. Fe thermochemical data was added from NIST-JANAF data \citep{Chase1998}. 

\section{Experimental Validation for Simplified Synthetic Systems}

The activity model for the liquid in VapoRock is that employed in the MELTS code \citep{Ghiorso1995}, which was itself calibrated on experimental data for basaltic liquids relevant to terrestrial magmatism. 
To better treat these complex systems, MELTS employs pseudo-components that simplify expressions for chemical potentials (and activities) of solutions. 
However, this approach comes at the cost of being able to access the entire space of liquid compositions. 
As such, the performance of the MELTS code in simpler ternary- and quaternary systems remains uncertain.

\begin{figure}[ht]
    \centering
    \includegraphics[width=0.45\textwidth]{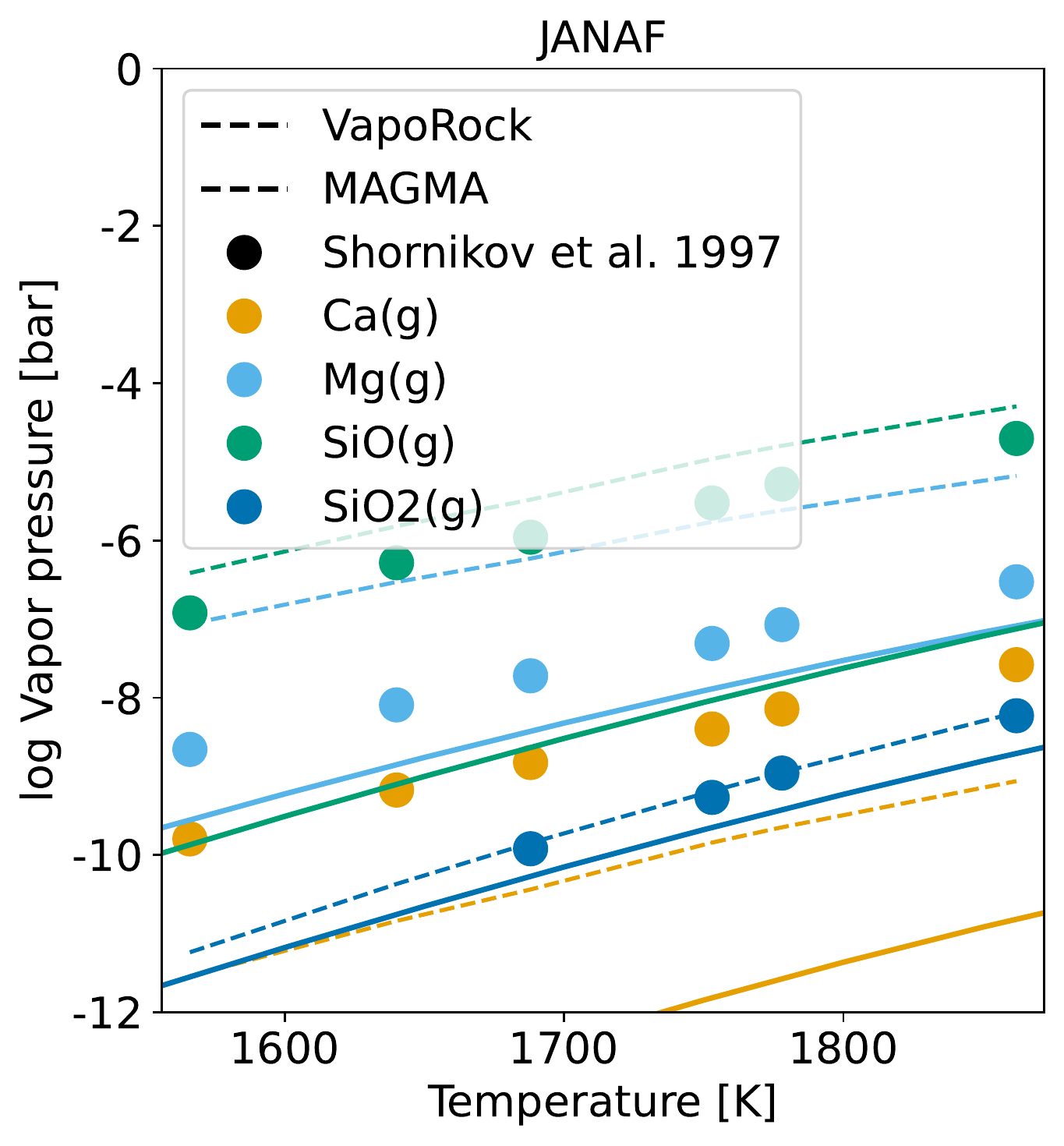} \includegraphics[width=0.45\textwidth]{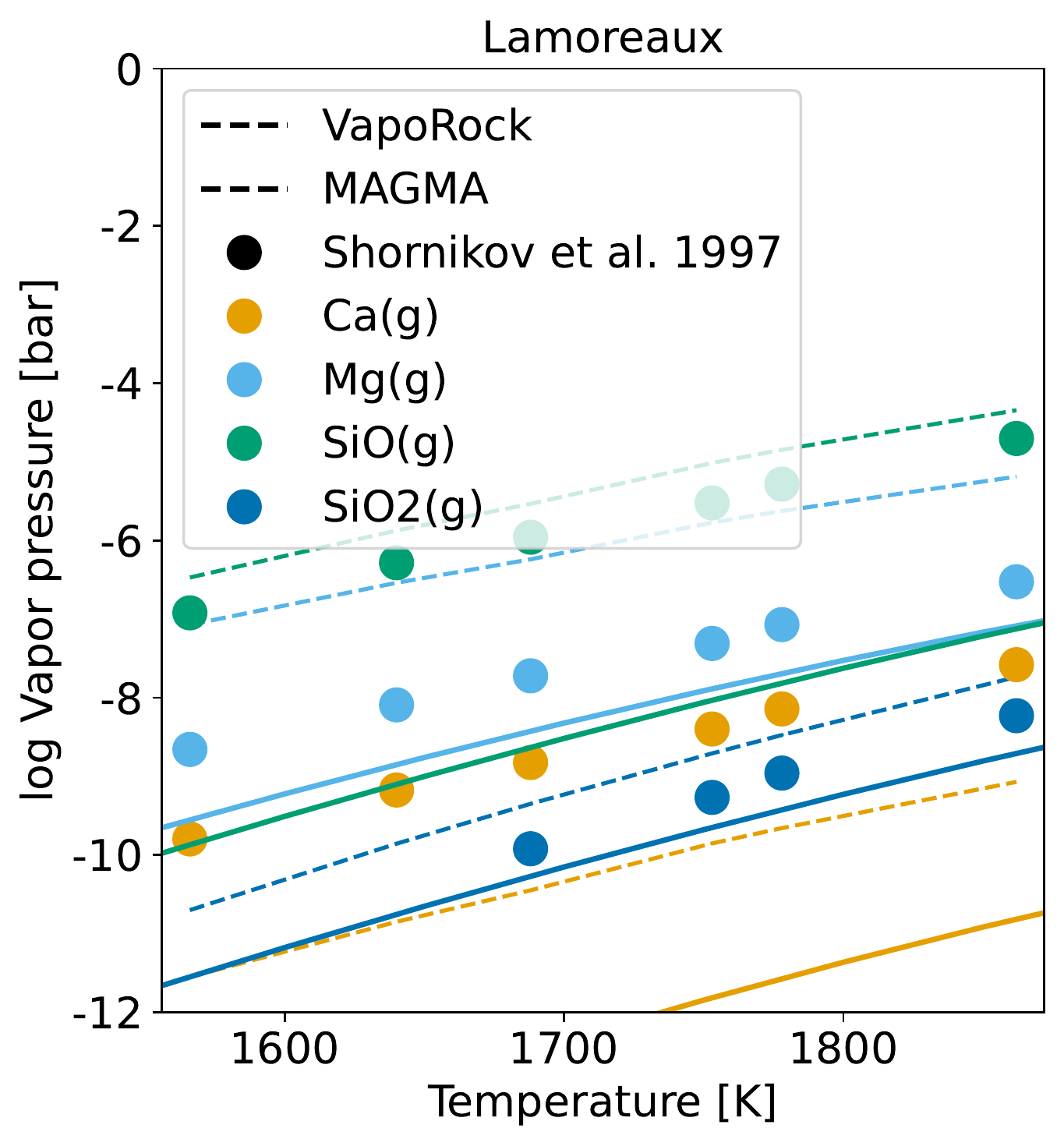}
    \caption{Partial pressures of vapor species determined by KEMS above diopside liquid (\cite{Shornikov1997}, points) compared with those calculated using VapoRock with either the JANAF \citep{Chase1998} or the Lamoreaux thermochemical data \citep[][]{Lamoreaux1984,Lamoreaux1987}. Although the partial pressures in some instances are in disaccord, the calculated activities from the vapor pressures shown coincide well with data from \cite{Rein1965} (Table~\ref{tab:simplesystems}). See text for discussion.}
    \label{fig:Diopside-validation}
\end{figure}

Given the underlying importance of the liquid activity coefficients in determining the vaporized species abundances, an additional validation method involves comparison of experimentally determined activity coefficients with those predicted by the model.
In these phase equilibria experiments, the compositions of two phases (a silicate liquid and a metal or sulfide liquid) are measured analytically. 
Provided that the thermodynamic properties for the metal or sulfide liquid (in terms of an activity-composition model for one or more of the components in each phase) are well-known, the activity of the unknown component can be determined thermodynamically, e.g. by integrating the Gibbs-Duhem equation \citep[e.g.][]{oneilleggins2002, holzheid1997, Rein1965, woodwade2013}. 
To examine the accuracy of both the vapor pressures and liquid activities predicted by VapoRock, we performed calculations at given temperature and oxygen fugacity to match those in experimental determinations of partial pressures and melt component activities in diopside \citep{Shornikov1997}, anorthite \citep{stolyarova1996}, the anorthite-diopside eutectic \citep{Rein1965}, and the system FeO-MgO-\ch{SiO2} \citep{Plante1992}. 
The results are compared in Figure~\ref{fig:Diopside-validation} and Table~\ref{tab:simplesystems}.

\begin{deluxetable*}{rlcDDDDD}[!htbp]
\renewcommand{\arraystretch}{1.3}
\caption{Comparison of activity coefficients determined in simple silicate liquids by a variety of methods with those returned from VapoRock.\label{tab:simplesystems}}
\tablehead{
\multicolumn{1}{l}{Composition} & Method & T [K] & \multicolumn{2}{c}{$\gamma$\ch{SiO2}} & \multicolumn{2}{c}{$\gamma$\ch{Al2O3}} & \multicolumn{2}{c}{$\gamma$MgO} & \multicolumn{2}{c}{$\gamma$CaO} & \multicolumn{2}{c}{$\gamma$FeO} \\
}
\decimals
\startdata
        \multicolumn{1}{l}{Diopside liquid} & ~ & ~ & ~ & ~ & ~ & ~ & ~ \\ \hline
        S$+$97  & KEMS          & 1873& 0.32 & - & 0.006 & 0.070 & - \\ 
        RC65    & Gibbs-Duhem   & 1873 & 0.50 & - & 0.090 & 0.004 & - \\ 
                & VapoRock      & 1873 & 0.59 & - & 0.105 & 0.002 & - \\
                & MAGMA 22      & 1873 & 0.17 & -  & 0.081 & 0.002 & - \\
        \multicolumn{1}{l}{Anorthite liquid }& ~ & ~ & ~ & ~ & ~ & ~ & ~ \\ \hline
        S$+$96  & KEMS          & 1933 & 0.52 & 0.69 & - & 0.0190 & - \\
                & VapoRock      & 1933 & 0.57 & 0.60 & - & 0.0014 & - \\
                & MAGMA 22      & 1933 & 0.40 & 0.27 & - & 0.0006 & - \\ 
        RC65    & Gibbs-Duhem   & 1823 & 0.42 & 0.80 & - & 0.0010 & - \\ 
                & VapoRock      & 1823 & 0.57 & 0.60 & - & 0.0009 & - \\
                & MAGMA 22      & 1823 & 0.42 & 0.24 & - & 0.0004 & - \\
        \multicolumn{1}{l}{An$_{42}$-Di$_{58}$} & ~ & ~ & ~ & ~ & ~ & ~ & ~ \\ \hline
        RC65    & Gibbs-Duhem   & 1873 & 0.45 & - & - & - & - \\ 
                & VapoRock      & 1873 & 0.64 & 0.23 & 0.133 & 0.0016 & - \\
                & MAGMA 22      & 1873 & 0.23 & 0.24 & 0.065 & 0.0011 & - \\
        \multicolumn{1}{l}{45S~:~45M~:~10F} & ~ & ~ & ~ & ~ & ~ & ~ & ~ \\ \hline
        P$+$92  & KEMS          & 1973& - & - & - & - & 4.04 \\ 
                & VapoRock      & 1973 & 1.13 & - & 0.104 & - & 2.48 \\
                & MAGMA 22      & 1973 & 0.31 & - & 0.079 & - & 0.83 \\
        \multicolumn{1}{l}{40S~:~40M~:~20F} & ~ & ~ & ~ & ~ & ~ & ~ & ~ \\ \hline
        P$+$92  & KEMS          & 1973 & - & - & - & -      & 3.02 \\ 
                & VapoRock      & 1973 & 1.16 & - & 0.112 & - & 1.95 \\
                & MAGMA 22      & 1973 & 0.28 & - & 0.093 & - & 0.77 \\
        \multicolumn{1}{l}{35S~:~35M~:~30F} & ~ & ~ & ~ & ~ & ~ & ~ & ~ \\ \hline
        P$+$92  & KEMS          & 1973 & -  & - & -     & - & 2.33 \\ 
                & VapoRock      & 1973 & 0.63 & - & 0.169   & - & 2.48 \\
                & MAGMA 22      & 1973 & 0.25 & - & 0.11 & - & 0.74 \\
\enddata
    \tablecomments{S~=~\ch{SiO2}, M~=~MgO, F~=~FeO, S+97~=~\cite{Shornikov1997}, RC65~=~\cite{Rein1965}, S+96~=~\cite{stolyarova1996}, P+92~=~\cite{Plante1992}.}
\end{deluxetable*}

The equilibrium partial pressures above diopside liquid as calculated by VapoRock agree favourably with those determined by KEMS \citep{Shornikov1997} for SiO(\textit{g}) and \ch{SiO2}(\textit{g}), particularly when considering the free energy data for gases from the JANAF tables (Fig.~\ref{fig:Diopside-validation}). 
The agreement, however, is considerably worse for Mg(\textit{g}) and Ca(\textit{g}), where both the JANAF- and Lamoreaux-based versions of VapoRock predict higher and lower partial pressures, respectively, by a factor $\sim$20 in each case. 
This systematic offset indicates that the activities of MgO(\textit{l}) and CaO(\textit{l}) in VapoRock are higher, and lower, respectively, than those derived from KEMS-determined partial pressures (see Table~\ref{tab:simplesystems}). 
However, when comparing the VapoRock activities to those determined by \cite{Rein1965}, the concordance is much improved, to within $\sim$15 \% for $\gamma$MgO and $\gamma$\ch{SiO2} and a factor of $\sim$2 for $\gamma$CaO.

For anorthite liquid at 1933 K, a similar discrepancy is observed for $\gamma$CaO between KEMS (0.019) and VapoRock (0.0014). 
The result is that VapoRock predicts lower partial pressures at a given temperature than the KEMS data indicate.
Again, however, the $\gamma$CaO is in excellent agreement with that determined by \cite{Rein1965}, as well as the model of \cite{berman1984}. 
The silica activity determined by VapoRock for the anorthite-diopside eutectic composition is also in good agreement with the data of \cite{Rein1965}.
Owing to the scarcity of experimental data, it is difficult to assess the relative accuracy of either dataset (KEMS vs. phase equilibria). 
However, the KEMS experiments of \cite{stolyarova1996, Shornikov1997} were performed in Mo Knudsen cells, which contributed significantly to the mass spectrum of the vapor phase, indicating Mo did not behave in an inert manner under their experimental conditions. 
Moreover, \cite{demaria1969} cited Mo as inappropriate  for use as Knudsen cell material owing to its reactivity with melt components and vapor species. 
Indeed, \cite{oneilleggins2002} noted a negative correlation between $\gamma$\ch{MoO3} and \textit{x}CaO in CMAS melts, which they attributed to the formation of \ch{CaMoO4} complexes. 
For these reasons, we would encourage additional investigation of these systems by KEMS using more suitable Knudsen cell material, namely Ir or Re \citep{Bischof2021, deMaria1971}.

In iron-bearing systems experimental determination of activities are complicated by its multiple oxidation states.
Nevertheless, the KEMS-measurements of \cite{Plante1992} examined liquids in the \ch{SiO2}-MgO-FeO system at molar \ch{SiO2}/MgO = 1 and varying \textit{x}FeO between 0.1 and 0.9. 
Direct comparison of vapour pressures was not performed, as \cite{Plante1992} used only measured ion intensities (e.g., \ch{Fe^+}) and by integrating the Gibbs-Duhem equation \citep[cf.][]{beltonfruehan1967} at constant \ch{SiO2}/MgO.
Due to the constrained compositional space of the MELTS model, VapoRock is unable to access compositions with too little silica, and thus comparison with results from \citet{Plante1992} is restricted to compositions with \textit{x}FeO $\le$ 0.3.
In each of the compositions tested (see predicted activities  in  Table~\ref{tab:simplesystems}), the value of $\gamma$FeO is within a factor of 1.5 of that determined by \cite{Plante1992}. 
Moreover, $\gamma$FeO has been widely determined in simple silicate melts in the CaO-MgO-\ch{SiO2}, CaO-\ch{Al2O3}-\ch{SiO2} and CaO-MgO-\ch{Al2O3}-\ch{SiO2} systems by the phase equilibria method, and is typically within the range 1 $\leq$ $\gamma$FeO $\leq$ 3 at similar temperatures \citep{holzheid1997, oneilleggins2002, woodwade2013}.
For each of these synthetic compositions, we also compare the predicted values using the MAGMA 22 code.
In general, while we find comparable predictions for $\gamma$MgO \& $\gamma$CaO, the values for $\gamma$\ch{SiO2}, $\gamma$\ch{Al2O3}, \& $\gamma$FeO are systematically lower than expected based on both experimental measurements and VapoRock calculations.
This exercise demonstrates the applicability of the MELTS code, even in simple ternary and quaternary systems (as long as they fall within the MELTS compositional bounds), to a typical level of accuracy of better than a factor of 2.

\bibliography{references.bib}{}